\documentclass[review,number,sort&compress]{elsarticle}

\usepackage{lineno}
\usepackage{hyperref}

\usepackage{graphicx}

\usepackage{amssymb}

\usepackage{commath}

\usepackage{epstopdf}

\usepackage{siunitx}
\journal{Nuclear Instruments and Methods A}

\begin{document}

\begin{frontmatter}

\title{Testbeam studies of a TORCH prototype detector}

\ead{mvandijk@cern.ch}

\address[A]{H.H. Wills Physics Laboratory, University of Bristol, Tyndall Avenue, Bristol BS8 1TL, United Kingdom}
\address[B]{European Organisation for Nuclear Research (CERN), CH--1211 Geneva 23, Switzerland}
\address[P]{Photek Ltd., 26 Castleham Road, St Leonards on Sea, East Sussex, TN38 9NS, United Kingdom.}
\address[C]{Denys Wilkinson Laboratory, University of Oxford, Keble Road, Oxford OX1 3RH, United Kingdom}

\fntext[O]{Now at University of Bath, Claverton Down, Bath BA2 7AY, United Kingdom.}
\fntext[Q]{Now at Universit\"at Gie\ss{}en, II. Physikalisches Institut, Heinrich-Buff-Ring 16, D--35392 Gie\ss{}en, Germany.}
\fntext[R]{Also at University of Leicester, Department of Physics and Astronomy, University Rd, Leicester LE1 7RH, United Kingdom.}
\fntext[S]{Now at Instituto de F\'isica Corpuscular, Calle Catedr\'atico Jos\'e Beltran, 2, 46980 Paterna, Valencia, Spain.}

\cortext[cor_auth]{Corresponding author.}

\author[A,O]{N.H.~Brook}
\author[B]{L.~Castillo~Garc\'ia}
\author[P]{T.M.~Conneely}
\author[A]{D.~Cussans}
\author[C,E]{M.W.U.~van~Dijk\corref{cor_auth}}
\author[B,Q]{K.~F\"ohl}
\author[B]{R.~Forty}
\author[B]{C.~Frei}
\author[C]{R.~Gao}
\author[B]{T.~Gys}
\author[C]{T.H.~Hancock}
\author[C]{N.~Harnew}
\author[P,R]{J.~Lapington}
\author[P]{J.~Milnes}
\author[B]{D.~Piedigrossi}
\author[C]{J.~Rademacker}
\author[A,S]{A.~Ros~Garc\'ia}

\begin{abstract}
TORCH is a novel time-of-flight detector that has been developed to provide charged-particle identification between 2 and 10~GeV/c momentum. TORCH combines arrival times from multiple Cherenkov photons produced within a 10~mm-thick quartz radiator plate, to achieve a 15~ps time-of-flight resolution per incident particle. A customised Micro-Channel Plate photomultiplier tube (MCP-PMT) and associated readout system utilises an innovative charge-sharing technique between adjacent pixels to obtain the necessary 70~ps time resolution of each Cherenkov photon. A five-year R\&D programme has been undertaken, culminating in the construction of a small-scale prototype TORCH module. In testbeams at CERN, this prototype operated successfully with customised electronics and readout system. A full analysis chain has been developed to reconstruct the data and to calibrate the detector. Results are compared to those using a commercial Planacon MCP-PMT, and single photon resolutions approaching 80~ps have been achieved. The photon counting efficiency was found to be in reasonable agreement with a GEANT4 Monte Carlo simulation of the detector. The small-scale demonstrator is a precursor to a full-scale TORCH module (with a radiator plate of $660\times1250\times10~{\rm mm^3}$), which is currently under construction.
\end{abstract}


\end{frontmatter}

\section{Introduction}
\label{section intro Introduction}

TORCH (Time Of internally Reflected CHerenkov light) is a novel detector\,\cite{Charles_2011_TORCH, Gys_2016_RICH_proceedings}, under development to provide time-of-flight (ToF) over a large-area, up to around 30\,m$^2$. The detector provides charged-particle identification between 2 and 10~GeV/c momentum over a flight distance of 10~m, and expands on the DIRC concept pioneered by the BaBar DIRC (Detection of Internally Reflected Cherenkov light) \cite{Adam_2005_DIRC-PID-for-BaBar} and the Belle-II iTOP \cite{Abe_2010_Belle-II-TDR} collaborations. TORCH combines fast timing information with DIRC-type reconstruction, aiming to achieve a ToF resolution of approximately 10-15\,ps per incident charged track. TORCH uses a thin 10\,mm quartz sheet as the radiator, utilizing the fast signal from prompt Cherenkov radiation. Total internal reflection is used to propagate the photons to the perimeter of the radiator, where they are focused onto an array of Micro-Channel Plate photomultiplier tube (MCP-PMT) photon detectors, which measure photon angles and arrival times. \\

The time difference between a pion and kaon over a 10\,m flight path is 35\,ps at 10\,GeV/$c$, therefore a per-track time resolution of 10--15\,ps is necessary to achieve a three sigma pion/kaon separation. This leads to a required single-photon time resolution of 70\,ps, given the expectation of about 30 detected Cherenkov photons per individual track. To attain this level of performance, simulation has shown that a 1\,mrad resolution is required on the measurement of the photon angle \,\cite{Charles_2011_TORCH}. To meet this requirement, MCP-PMTs of 53$\times$53 mm$^2$ active area and pixel granularity 128$\times$8 are necessary. Such detectors have been custom-developed for the TORCH application by an industrial partner, Photek Ltd.\\

A five-year R\&D programme has been undertaken, culminating in the design and construction of a small-scale prototype TORCH module. This module consists of a quartz plate of dimensions 120\,mm width, 350\,mm length, and 10\,mm thickness, read out by a single customised MCP-PMT of 32$\times$32 pixels filling a 26.5$\times$26.5 mm square, a quarter area of the final tube dimensions. The prototype was tested at the CERN Proton Synchrotron in 2015 and 2016 in a 5~GeV/$c$ pion/proton mixed beam, and the results compared to those measured with a commercial Planacon MCP-PMT. As a result of the testbeam studies, the full functionality of the TORCH design and its timing properties have been verified. \\

The small-scale demonstrator is a precursor to a full-scale TORCH module (660$\times$1250$\times$10~mm$^3$), read out by ten MCP-PMTs, which is currently under construction. The TORCH detector has been proposed to complement the particle identification of the LHCb experiment at CERN for its future upgrade\,\cite{LHCb_2011_LOI_upgrade, LHCb_2017_EOI_upgrade}.\\

In this paper, first an overview is given of the design of the TORCH detector and the principles of photon reconstruction. The optical system, the MCP-PMT detectors and the electronics readout system are described. The testbeam setup is then discussed, as well as the operating configurations. The method of data analysis, the algorithms for reconstruction, pattern recognition and data calibration are detailed. The single-photon time resolution is presented and the photon detection efficiency is compared to Monte Carlo expectations. Finally, a summary and an overview of future work is given.

\section{Design of TORCH}
\label{section Design of TORCH}

The TORCH detector concept involves precision timing of Cherenkov photons that are emitted by a charged particle as it passes through a solid radiator. The chosen radiator material is synthetic fused silica due to its suitable refractive index, good transmission, and radiation resistance. The radiator takes the form of a highly polished plate, with nominal thickness of 10~mm, chosen as a compromise between providing sufficient yield of detected photons and limiting the material budget of the detector. A large fraction of the photons generated are trapped within the plate by total internal reflection, and propagate to the edges where they are detected with a focusing system equipped with finely-pixellated fast photodetectors. These are located in the focal plane of the cylindrical focusing optics. After correcting for the time of propagation of each photon in the optical system, the photon provides a measurement of the time at which the particle crossed the plate. By combining the information from the different photons emitted by the particle, a high precision measurement can be made of its arrival time. In order to achieve an overall resolution of 70~ps per photon, a time resolution of 50~ps per photon from the photodetector (including the associated readout electronics) is needed, with a similar precision from the reconstruction of the time of propagation. With 30 detected photons from each charged particle, this would provide a timing resolution per particle of 15~ps, assuming the individual photon measurements are independent.\\

\begin{figure}
    \begin{center}
    \includegraphics[width=0.99\columnwidth]{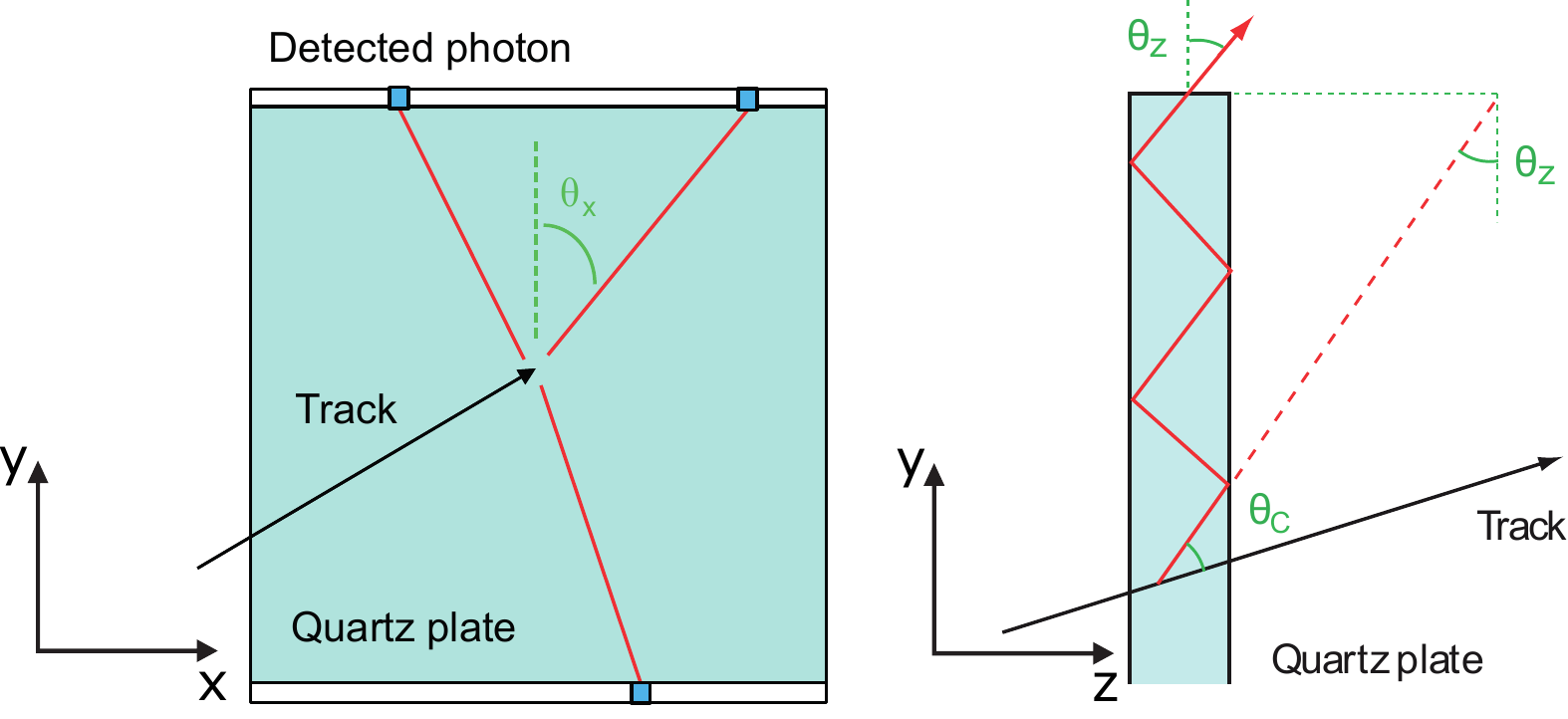}
        \caption{Schematic view of the TORCH radiator plate from the front ($xy$, left) and side ($yz$, right); the angles $\theta_x$ and $\theta_z$ are defined for a Cherenkov photon generated at an angle $\theta_C$ from a charged particle track \cite{Charles_2011_TORCH}.}
         \label{figure intro TORCH thx and thz definition}
    \end{center}
\end{figure}

\begin{figure}[!hbt]
    \centerline{\includegraphics[width=0.7\columnwidth]{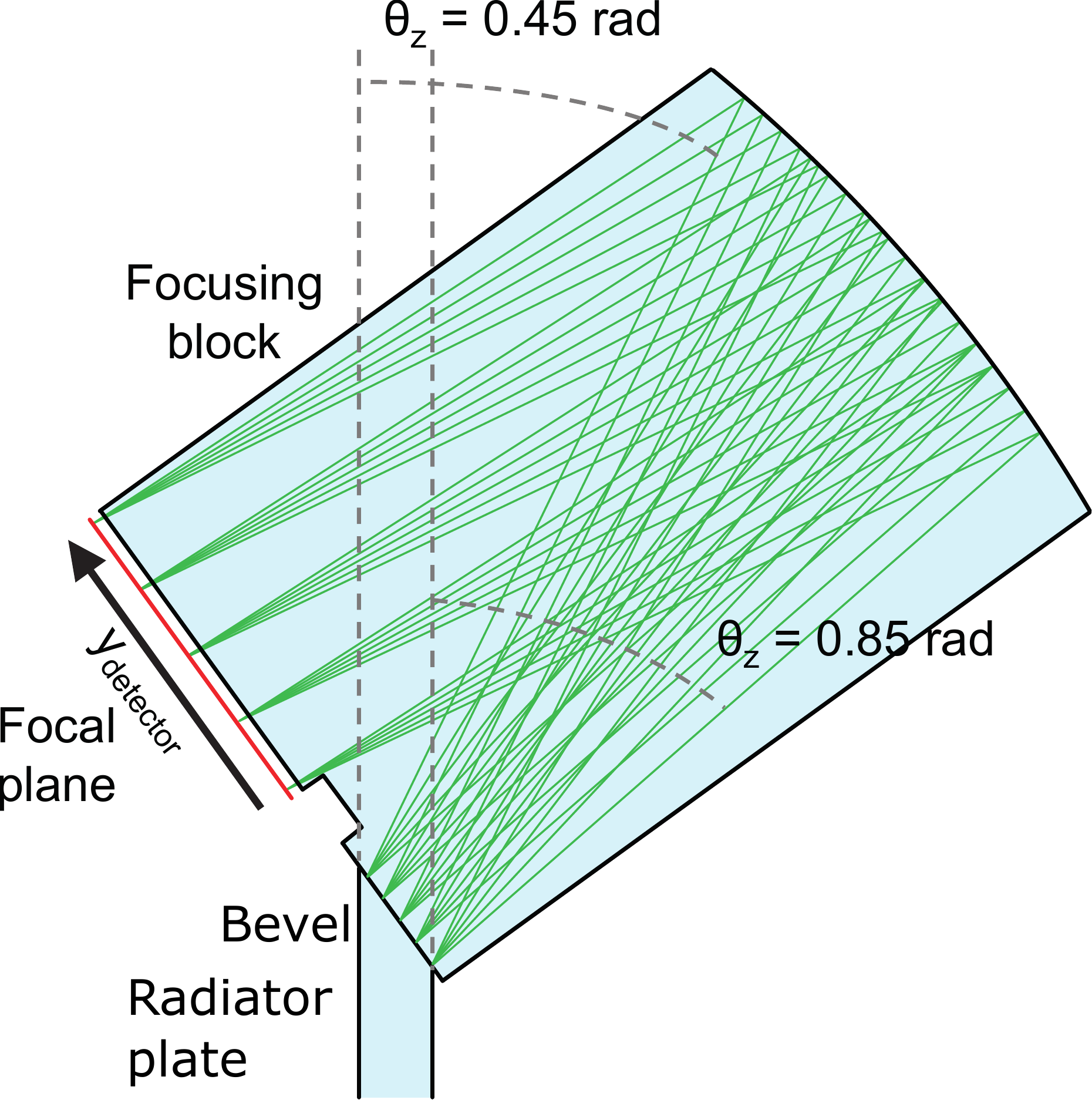}}
    \caption{Cross-section of the TORCH focusing block \cite{Castillo-Garcia_2014_NDIP-proceedings}; the right-hand side is a cylindrical mirror, and the paths of photons entering at the accepted range of angles are shown. The vertical detector coordinate is oriented along the focal plane. }
    \label{figure intro focusing block}
\end{figure}

Provided that the impact point of the charged particle on the radiator is known, the position of the detected photon along the coordinate $x$ at the plate top or bottom surface (see Figure \ref{figure intro TORCH thx and thz definition}) can provide a precise determination of the photon angle of propagation in the plate, $\theta_x$.  The angle of the photon in the second projection $\theta_z$ is determined using a focusing system \cite{Castillo-Garcia_2014_NDIP-proceedings}, which takes the form of a block of synthetic fused silica with a cylindrical mirrored surface, shown in Figure \ref{figure intro focusing block}. This converts the photon angle $\theta_z$ into a position along the local $y_{detector}$ axis, defined in Figure \ref{figure intro focusing block}, which is hereon referred to as the {\it vertical} coordinate of the photon detector (i.e. the focussing coordinate). Similarly the local $x_{detector}$ axis, which lies also along the $x$ axis, is referred to as the {\it horizontal} (non-focussing) coordinate of the photon detector. Monte Carlo studies have shown a precision of about 1~mrad is required on the angle of the photon in both $\theta_x$ and $\theta_z$ projections, to achieve the required resolution on the time of propagation of the photon as it reflects within the plate \cite{Charles_2011_TORCH}. \\

The largest commercially available size of MCP-PMTs with proven technology, the Photonis Planacon, is 60$\times$60~mm$^2$ with an active area of 53$\times$53~mm$^2$ \cite{Photonis_2014_Planacon_datasheet}. The lower limit on $\theta_z$ of 0.45 rad is set by the largest vertical track angle (about 250 mrad, for a 2.5~m high radiator at 10~m distance) plus the largest Cherenkov angle for which light can be detected (about 900 mrad at 7~eV photon energy), generated by a track which is undeviated from the interaction point. The upper limit on $\theta_z$ of 0.85 rad is set by the smallest angle that will still give total internal reflection. In the vertical detector direction, dividing this 400~mrad range into 128~pixels allows for the 1~mrad requirement on $\theta_z$ to be achieved, given that the resolution of a pixel scales with the pixel size as $1/\sqrt{12}$. In $\theta_x$, assuming the Cherenkov photons can propagate at least 2~m after generation, the 1~mrad requirement is met by employing 8~pixels per detector in the horizontal direction. This gives the final design requirement on the effective pixel size to be 6.625$\times$0.414~${\rm mm^2}.$\\

Figure \ref{figure intro path length calculation} shows an illustration of the photon path length calculation, by unfolding the multiple reflections the photon undergoes. The path length $L$ can be calculated by projecting the initial direction vector of the photon over the difference in height between track impact point and photon detector, $H$. $L$ is then given by the geometrical projection:

\begin{figure}[!hbt]
    \centerline{\includegraphics[width=0.7\columnwidth]{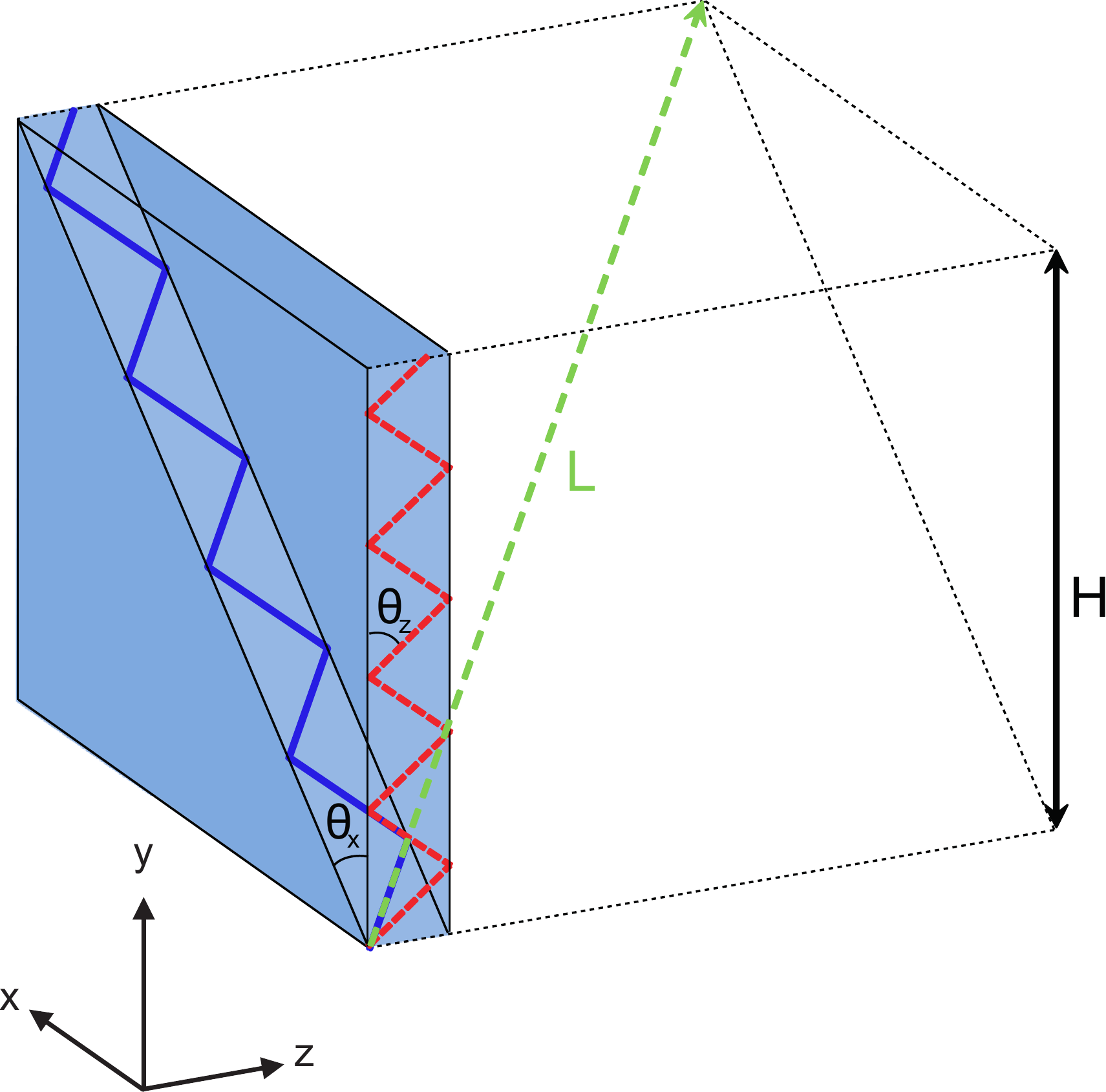}}
    \caption{Illustration of the photon path length calculation: the photon path is shown as the blue line within the radiator plate (shaded), which is unfolded into the green dashed line.}
    \label{figure intro path length calculation}
\end{figure}

\begin{equation}
	L = \sqrt{\frac{H^2}{\cos^2{\theta_x}} + \frac{H^2}{\cos^2{\theta_z}}-H^2}\quad .
	\label{equation intro path length calculation}
\end{equation}

Due to chromatic dispersion in the radiator, photons with different energies $E$ propagate at different speeds, which needs to be corrected for. The speed is governed by the group refractive index $n_g$, which can be derived from the phase refractive index $n_p$ of the material:

\begin{equation}
    \label{equation intro nphase ngroup}
    n_g = n_p + E \frac{{\rm d} n_p}{{\rm d} E}\quad .
\end{equation}

\begin{figure}[!hbt]
    \centerline{\includegraphics[width=0.7\columnwidth]{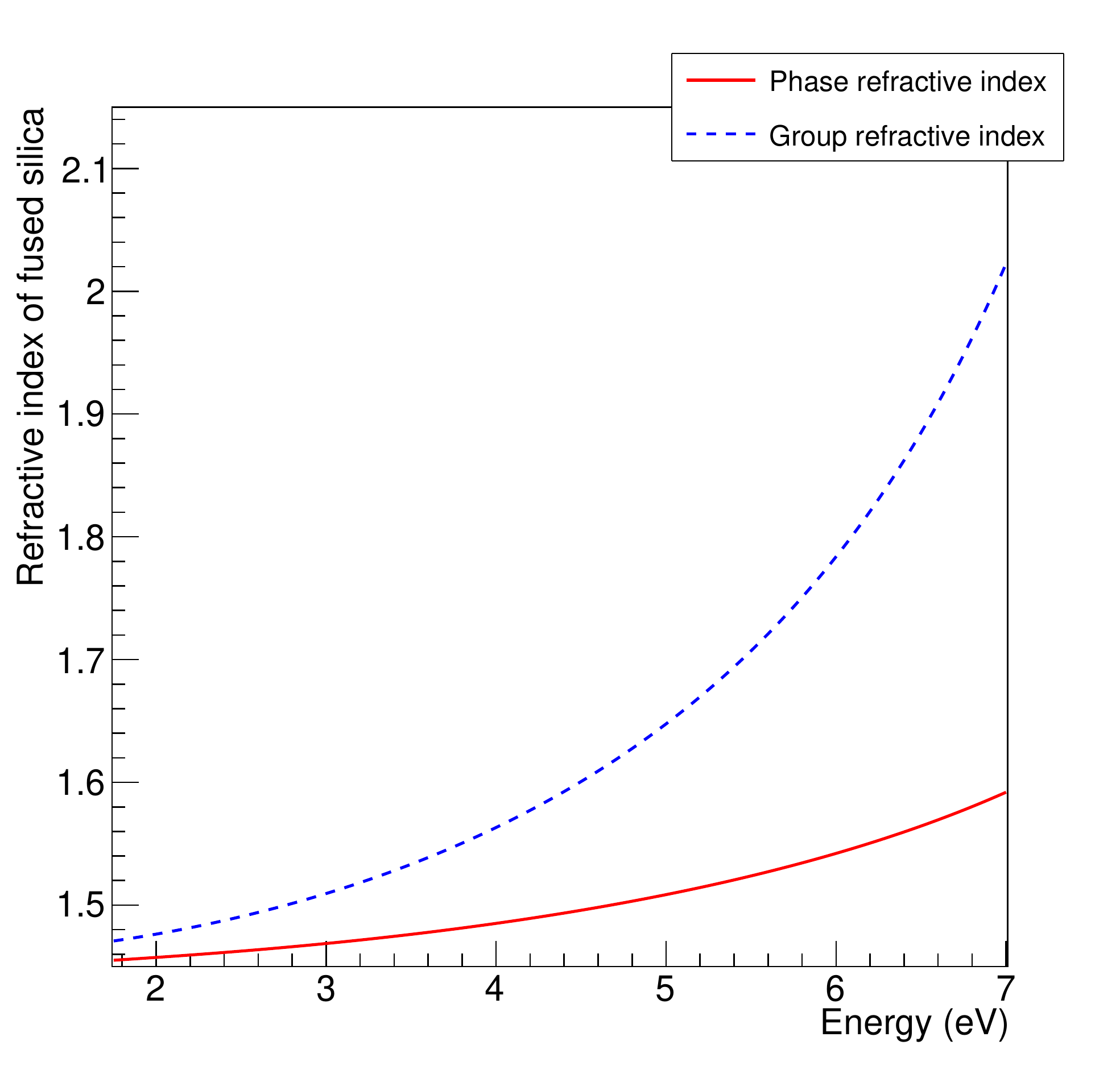}}
    \caption{Phase and group refractive indices of synthetic fused silica as a function of photon energy, as calculated from Ref.~\cite{Corning_2014_HPFS_7980}.}
    \label{figure intro refractive indices}
\end{figure}

\noindent The phase and group refractive indices for fused silica as a function of photon energy \cite{Corning_2014_HPFS_7980} are shown in Figure \ref{figure intro refractive indices}.  \\

The measured Cherenkov emission angle $\theta_C$ is used to correct for chromatic dispersion. Given the track and photon unit direction vectors $\mathbf{\hat{v}_t}$ and $\mathbf{\hat{v}_p}$:

\begin{equation}
	\mathbf{\hat{v}_t} \cdot \mathbf{\hat{v}_p} =
	\cos\, \theta_C = \frac{1}{n_p\,\beta}\quad ,
\label{equation intro cherenkov reconstruction}
\end{equation}

\noindent which allows $n_p$ to be determined. Then the group refractive index $n_g$ can be calculated using Eq.~\ref{equation intro nphase ngroup} and the known dispersion relations. This derivation of the refractive index also requires knowledge of $\beta$, the speed of the charged particle expressed as a fraction of the speed of light. Assuming the particle momentum is measured, $\beta$ can be calculated for each particle mass in turn, and propagated through the subsequent analysis, allowing the preferred mass hypothesis to be selected.\\

\begin{figure}[!hbt]
    \centerline{\includegraphics[width=0.99\columnwidth]{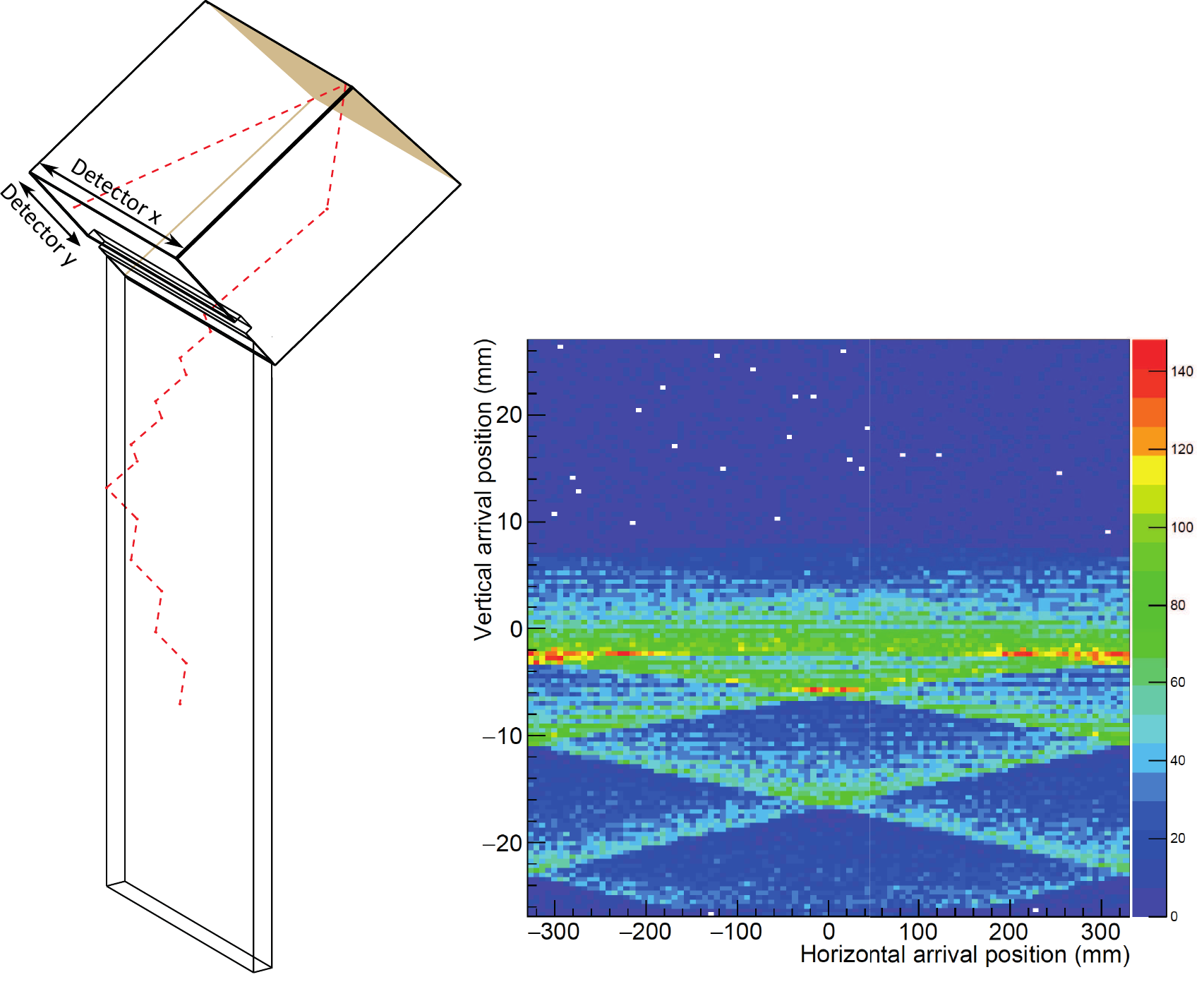}}
    \caption{(Left) Isometric view of a TORCH module of reduced size showing the path of a single photon propagating through the optics (red dashed line) and showing the definition of horizontal ($x_{detector}$) and vertical direction ($y_{detector}$); (Right) the folded pattern of photons expected to arrive at the photodetector plane for a full-sized TORCH module, derived from GEANT simulation.}
    \label{figure intro double side reflection}
\end{figure}

There are several contributions to the path length of each photon that need to be taken into account, namely the effects of a bevel at the top of the radiator plate, the focusing block and the photon detector window. The bevel (visible in Figure~\ref{figure intro focusing block}) is introduced to simplify the construction of the focusing block, but also adds an ambiguity in the photon path, since light propagating towards negative $z$ at this point will undergo an extra reflection off the front surface of the radiator plate. In addition, for practical reasons it is not feasible to make the full TORCH detector from a single radiator plate, and instead the detector will be subdivided into modules in the $x$ direction. For a large plate, the mapping of the Cherenkov cone through the optics gives rise to a roughly hyperbolic pattern of photon hits on the detectors but, when the plate is subdivided into modules, this pattern is folded on itself due to reflections at the vertical sides. This is illustrated in Figure~\ref{figure intro double side reflection} (left), where the path of a single photon is shown schematically, reflecting twice off a vertical side: once in the radiator plate and once in the focusing block. The folded pattern at the photodetector plane is shown in Figure \ref{figure intro double side reflection} (right) for a module of 66~cm wide with a radiator height of 2.5~m, which corresponds to the dimensions of a module in the final layout. The reflections at the module sides introduce ambiguities in the reconstructed path length. Whilst for a sufficiently wide module there are several solutions for the path that are consistent with a physical solution for the Cherenkov angle, the ambiguities can be resolved in the reconstruction.\\

The requirements for the optics, photon detector and readout electronics are now discussed in turn.

\subsection{Optics requirements}
\label{subsection Requirement on optics}

The yield of detected photons in the TORCH detector is limited by the optical components in two ways: scattering from surface roughness and Rayleigh scattering. Rayleigh scattering is a fundamental property that cannot be avoided. A Rayleigh scattering length of 500~m is assumed at an energy of 2.805~eV \cite{Cohen-Tanugi_2003_BABAR_DIRC_optical_properties}, scaling with the photon energy as $E^4$. \\

In order to limit losses from surface roughness, it is required that the large flat plate surfaces are polished to a roughness of less than 0.5~nm. Assuming this surface roughness, simulations show that about 14\% of the total number of photons that would otherwise propagate to the detector are lost \cite{Van_Dijk_2016_Thesis}. If this parameter is relaxed to 1~nm, the expected losses increase to about 32\%. \\

For manufacturing reasons the focusing optics and radiator plate are produced independently, and need to be optically coupled. Candidate glues have been tested~\cite{Castillo_Garcia_2016_thesis, Castillo-Garcia_2014_NDIP-proceedings}, including Epotek~301-2, Epotek~305 and Pactan~8030.  Epotek 301-2 was used in the BaBar DIRC \cite{Adam_2005_DIRC-PID-for-BaBar} and was found to be mechanically strong, stable and radiation hard, with a well-known refractive index~\cite{Montecchi_2001_CMS-glue-paper}. However, its transmission cuts off at a photon energy of about 4~eV.  Epotek~305 transmits up to about 5~eV, is appropriately radiation hard~\cite{Shetter_1979_Epotek-305-radiation-hardness} but limited information is available on its refractive index~\cite{Light_1978_Epotek-305-refractive-index}. Pactan~8030 is a silicone-based adhesive with even better transmission characteristics, up to 6~eV, although little information is available on its other physical properties. Unlike the other epoxy-based glues, Pactan~8030 allows for disassembly because it does not set rigidly. It is therefore suitable for the prototyping phase and has been used here.\\

\begin{figure}[!hbt]
    \centerline{\includegraphics[width=0.8\columnwidth]{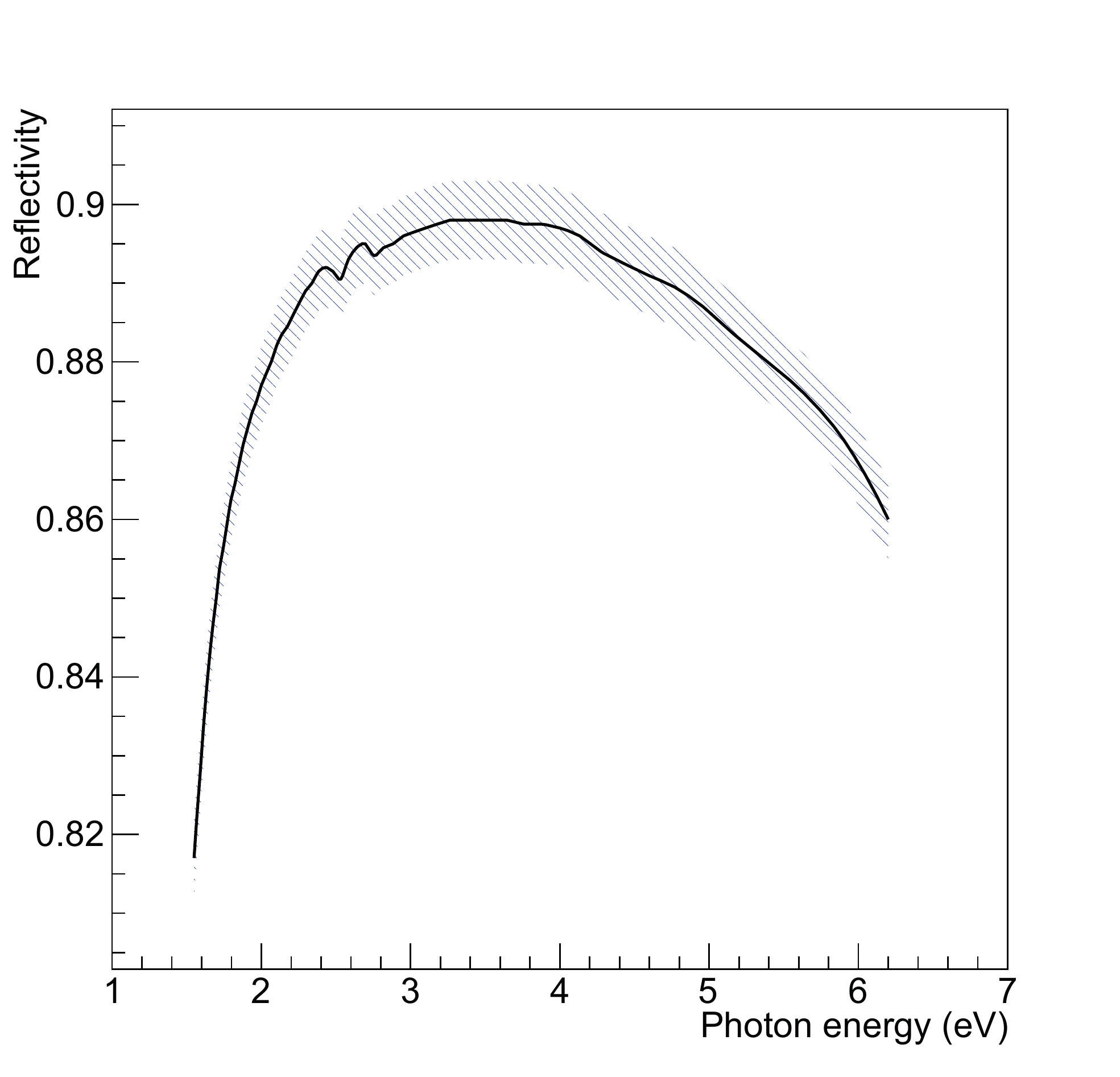}}
    \caption{Reflectivity of the mirror surface of the focusing block as a function of photon energy, measured on a 1~mm thick sample of Suprasil fused silica coated with $\sim$120~nm of aluminium, at an angle of incidence of 30 degrees. The measurement error is estimated at 0.5\% (absolute), indicated by the shaded band. }
    \label{figure intro mirror reflectivity}
\end{figure}

A further loss of photons in the TORCH optics comes from imperfect reflectivity of the cylindrical mirrored surface of the focusing block. The reflectivity of a representative aluminized sample has been measured, and is shown in Figure~\ref{figure intro mirror reflectivity}. The reflectivity is typically above 85\% for the photon energy range of interest.

\subsection{TORCH Photodetectors}
\label{subsection TORCH photodetectors}

To meet the TORCH requirements, the photodetectors require a very good intrinsic time resolution (20-30~ps), low dark noise, spatial granularity (at the anode), along with a high active to dead area ratio. Micro-Channel Plate photomultiplier (MCP-PMT) technology (for review, see \cite{Gys_2015_MCP_overview}) meets these requirements and has also been adopted by other DIRC-type detectors \cite{Abe_2010_Belle-II-TDR,Cowie_2009_PANDA-Disc-DIRC}. The main drawbacks are the relatively low detection efficiency compared to alternative technologies, a limitation on the lifetime, and the restricted granularity of commercial devices. \\

To address the issues of lifetime and granularity, a three-phase R\&D programme was instigated with an industrial partner, Photek Ltd. The first phase addressed the lifetime issues of the MCP-PMT on a small, circular MCP-PMT device (25~mm diameter) \cite{Gys_2016_RICH_proceedings}. The second phase demonstrated the granularity required for TORCH, implementing a square pixellated anode in a circular MCP-PMT device (40~mm diameter), which was used for the testbeam in 2015. The third phase\footnote{These final tubes were delivered in Summer 2017 and are currently undergoing tests.} combines all requirements in a square 60$\times$60~mm$^2$ MCP-PMT with a sensitive area of 53$\times$53~mm$^2$. The testbeam programmes in 2015 and 2016 considered both a Photek Phase-2 tube with an S20 multi-alkali photocathode and a commercially available tube from Photonis, the XP85122 \cite{Photonis_2014_Planacon_datasheet} with a bi-alkali photocathode. Both these MCP-PMTs employ micro-channel plates with a pore size of 10~$\mu$m. \\

The Photek PMTs have a double set of MCPs in a ``Chevron'' configuration. Photoelectrons can reflect off the front face of the first MCP giving rise to secondary signals; typically referred to as backscattering \cite{Korpar_2008_MCP-timing-crosstalk}. The signals from these photoelectrons are translated in space and arrive later in time, with the typical spread in translation and delay set by the distance between the photocathode and the first MCP. \\

\begin{figure}[!hbt]
\centerline{
\includegraphics[width=0.95\columnwidth]{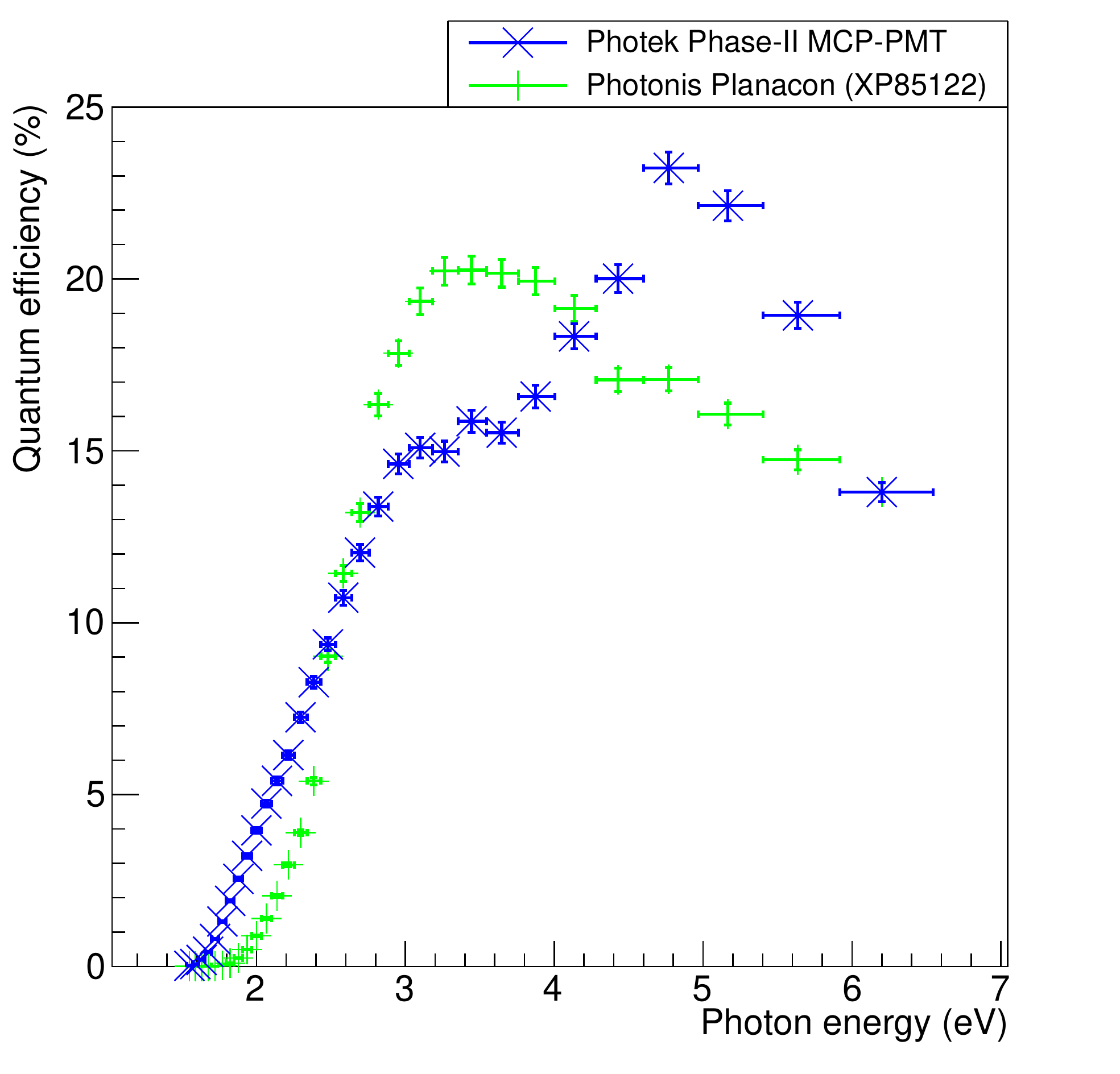}}
\caption{Measured quantum efficiency for a representative Photek Phase-2 tube featuring a multi-alkali S20 photocathode and a Photonis Planacon (XP85122) featuring a bi-alkali photocathode.}
\label{figure intro QE curves}
\end{figure}

The photon counting efficiency of the photodetector is determined by the collection efficiency and the quantum efficiency. The collection efficiency, here estimated to be 65\%, is defined as the ratio of generated to detected photoelectrons after photon conversion. The quantum efficiency is the ratio of photons incident on the front face of the photocathode and those that generate photoelectrons, and is highly dependent on the incident photon energy. The measured quantum efficiency of the two deployed phototube types is shown in Figure \ref{figure intro QE curves}. \\

During the development program it became apparent that, with the 8$\times$128 pixel requirement in a 53$\times$53~mm$^2$ active area, it would be difficult to fit this number of pixels within the envelope of the detector in the vertical direction. This was solved by halving the number of pixels to 64 and sharing the collected charge over multiple pixels. The electronics (further detailed in section \ref{subsection Electronics for TORCH}) perform a simultaneous charge and timing measurement for the PMT signal, and this information can be used in a charge-weighting algorithm to achieve a resolution significantly better than would be expected based on pixel size \cite{Conneely_2015_PSD_proceedings}. \\

To take advantage of the charge-sharing method, a novel technology was developed, which combined aspects from direct and capacitively coupled readout of the Photek tube~ \cite{Conneely_2015_PSD_proceedings}. The readout pads are directly coupled to electrodes buried beneath a thin dielectric layer. These electrodes pick up the charge induced by the electron shower emanating from the MCP stack and collected on the anode resistive layer. Varying the thickness of this resistive layer allows the degree of charge sharing between the pixels to be tuned. \\

The size of the electron shower generated in the MCP-PMT is governed by the tube electrostatics, the layout of the MCP stack and the gap from the rear of the MCP stack to the anode. From earlier measurements of the Photek tube \cite{Van_Dijk_2016_Thesis, Castillo_Garcia_2016_thesis} it is known that the size of the avalanche generated by the MCP-PMT is large compared to an individual pixel. Combined with the charge sharing between pixel pads it is expected that each single photon will have about 3--4 pixel hits. For the Phase-2 generation of MCP-PMTs, the MCP-anode gap is nominally 4.5~mm, and the thickness of the dielectric layer burying the anode contact pads is 0.5~mm.

\subsection{TORCH electronics}
\label{subsection Electronics for TORCH}
The electronics readout system is a key component in achieving the timing resolution required for the TORCH ToF measurement and has gone through an extensive programme of development\,\cite{Gao_2014_TWEPP2013_proceedings, Gao_2015_TWEPP2014_proceedings, Gao_2016_TWEPP2015_proceedings}. The readout to digitize the signals from the MCP-PMT is based on the NINO \cite{Anghinolfi_2004_NINO-for-ALICE} and the HPTDC \cite{Akindinov_2004_HPTDC-for-ALICE} chip-sets, both employed by the ALICE experiment. The NINO ASIC was originally developed as an 8-channel device, with the later 32-channel version \cite{Despeisse_2011_NINO32} utilized for TORCH. A front-end PCB containing two NINO chips reads out 64 MCP-PMT channels, which then connects into a second PCB containing two HPTDC chips, each of which operates as a 32-channel device with 97.7\,ps time binning. The NINO provides discrimination and amplification and takes as input a signal from the MCP-PMT and converts it into an LVDS output pulse, the width of which is a measure of the amount of charge in the signal. The HPTDC then digitises the LVDS pulse by time-stamping the leading and falling edges. This combination of ASICs gives rise to several calibrations which need to be performed in order to reach optimal timing performance:

\begin{itemize}
\item The charge-to-width calibration of the NINO;
\item A time-walk correction of the NINO leading edge (since this is a single-threshold discrimination device);
\item An Integral Non-Linearity (INL) correction to the HPTDC, which is a well documented feature\,\cite{Schambach_2003_HPTDC_at_STAR}.
\end{itemize}

These calibrations and their impact will be discussed in Section \ref{section Calibrations}.

\section{The TORCH prototype}
\label{section The TORCH prototype}

A small-scale TORCH prototype has been constructed featuring optical components of reduced size. Specialized mechanics have been produced for mounting the MCP-PMTs and the accompanying electronics. This prototype was constructed to demonstrate the feasibility of the TORCH concept and to determine the performance of the components used.

\subsection{Prototype optics}
\label{subsection Prototype optics}

\begin{figure}
    \begin{center}
        \begin{minipage}[!hbt]{0.563\columnwidth}
            \includegraphics[width=\textwidth]{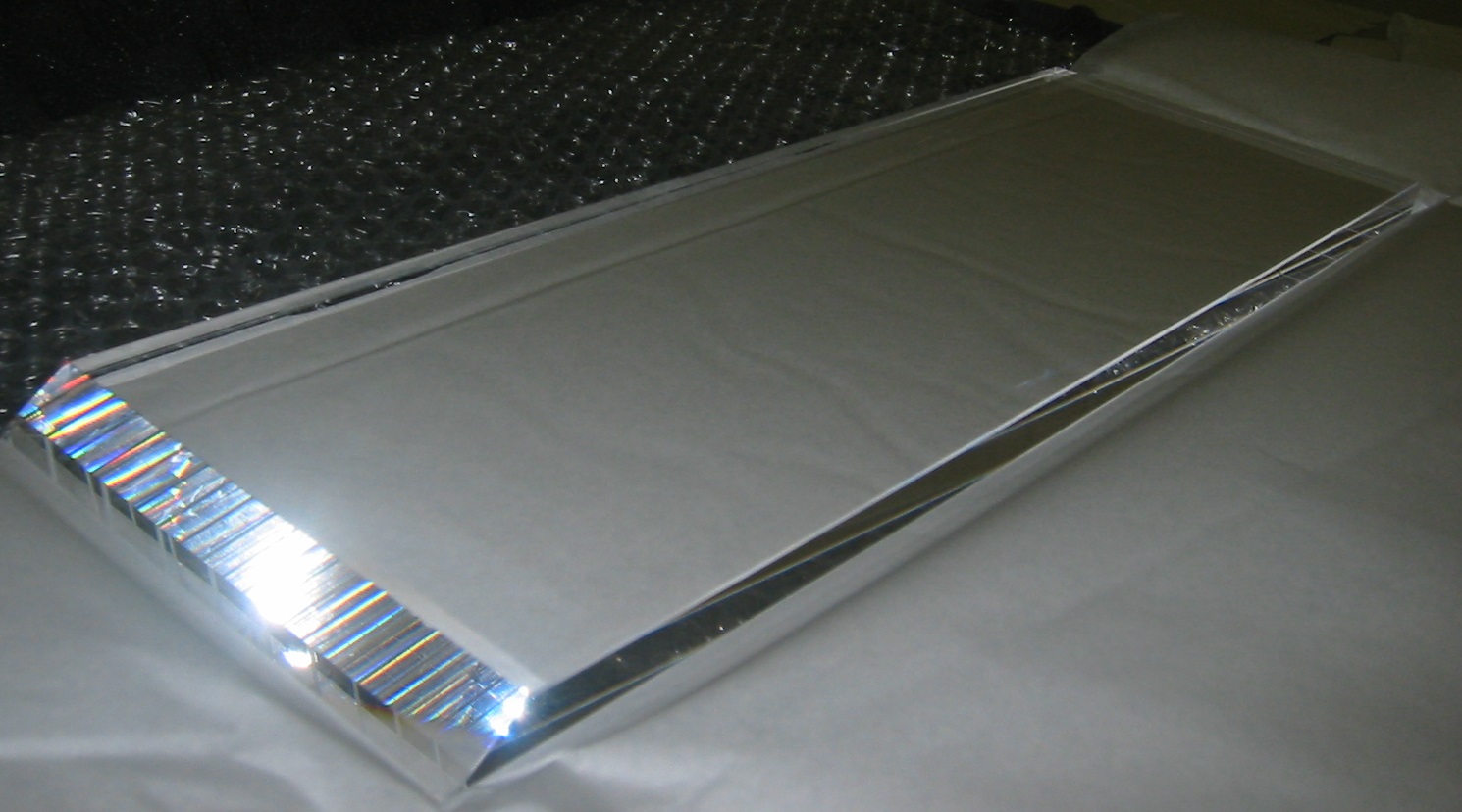}
        \end{minipage}
        \hspace{0.1cm}
        \begin{minipage}[!hbt]{0.387\columnwidth}
            \includegraphics[width=\textwidth]{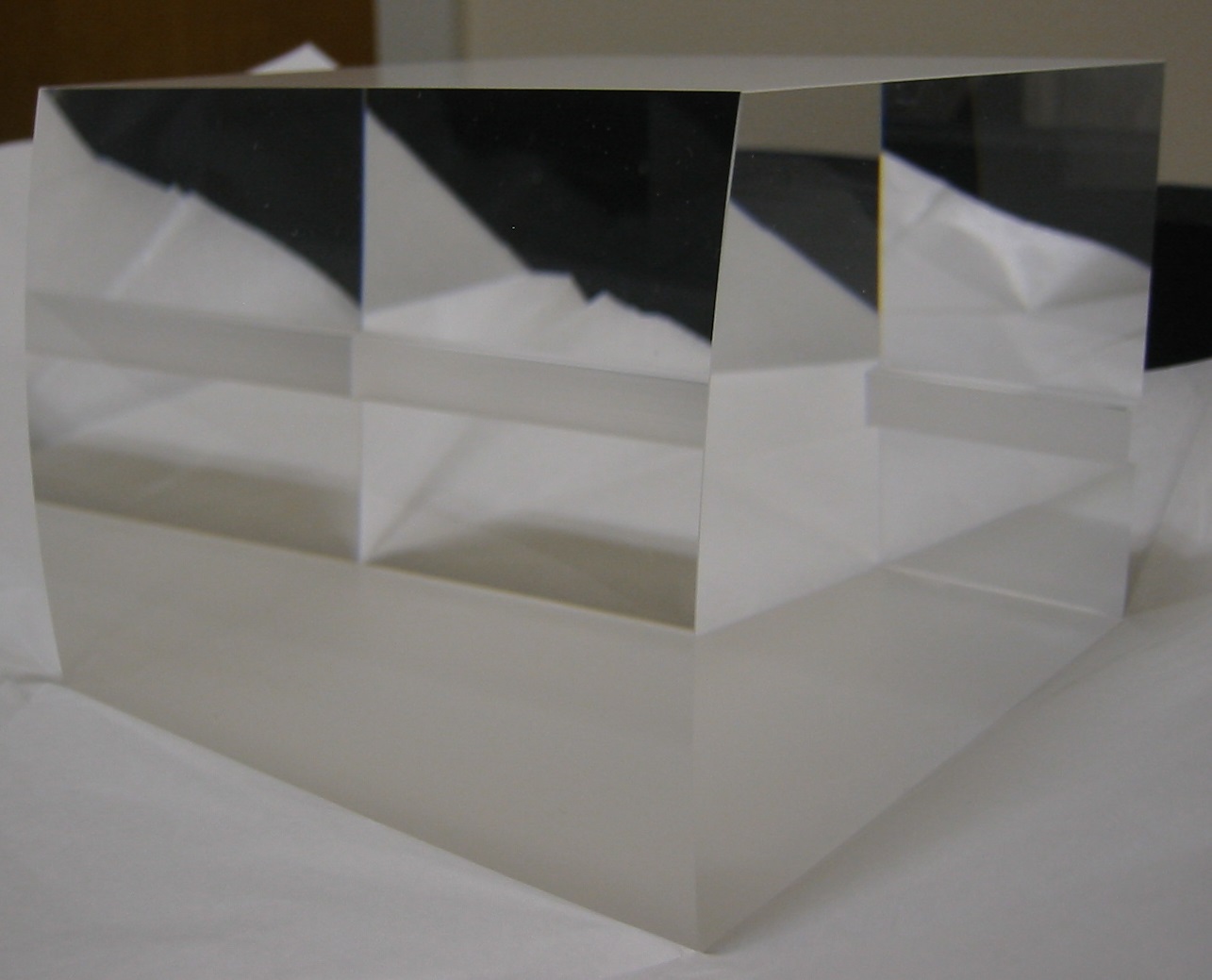}
        \end{minipage}
        \caption{Optical components procured from Schott for the TORCH testbeam prototype. (Left) The radiator plate of size 120$\times$350$\times$10~mm$^3$ showing the bevelled edge, of which the acute angle is 36$^\circ$. (Right) Matching focusing block of width 120~mm, with a cylindrical surface with focal length 260~mm, following the design shown in Figure \ref{figure intro focusing block}.}
        \label{figure PS2015 focusing optics}
        \end{center}
\end{figure}

The optical components of the TORCH prototype were produced from fused silica (specifically, Corning~7980) by Schott\footnote{SCHOTT Schweiz AG, St. Josefen-Strasse 20, 9001 St. Gallen, Switzerland.}. These components followed the design outlined in Section \ref{section Design of TORCH}, but with scaled-down dimensions: a radiator plate of 120$\times$350$\times$10~mm$^3$ (width$\times$height$\times$thickness) and focussing optics of matching width. The radiator plate was polished to a surface roughness of about 1.5~nm. While this is significantly less stringent than the requirement placed on the full-sized radiator plate (0.5~nm), the number of reflections that individual photons undergo is reduced due to the smaller size of the radiator, hence the requirement was relaxed on cost consideration. Photographs of both components are shown in Figure \ref{figure PS2015 focusing optics}. \\

The focusing block was manufactured to focus 2~mm beyond the exit surface onto the photocathode of the detector. The block was aluminized (see Figure \ref{figure intro mirror reflectivity}), and the quartz components glued together using Pactan--8030.

\subsection{MCP-PMTs and electronics}
\label{subsection MCP-PMTs and electronics}

\begin{figure}
    \begin{center}
        \begin{minipage}[!hbt]{0.5074\columnwidth}
            \includegraphics[width=\textwidth]{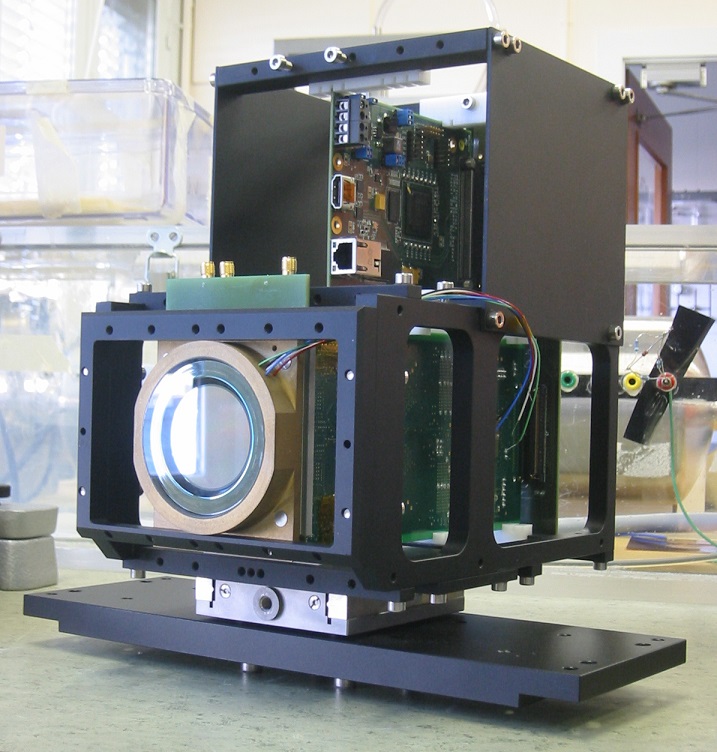}
        \end{minipage}
        \hspace{0.1cm}
        \begin{minipage}[!hbt]{0.4425\columnwidth}
            \includegraphics[width=\textwidth]{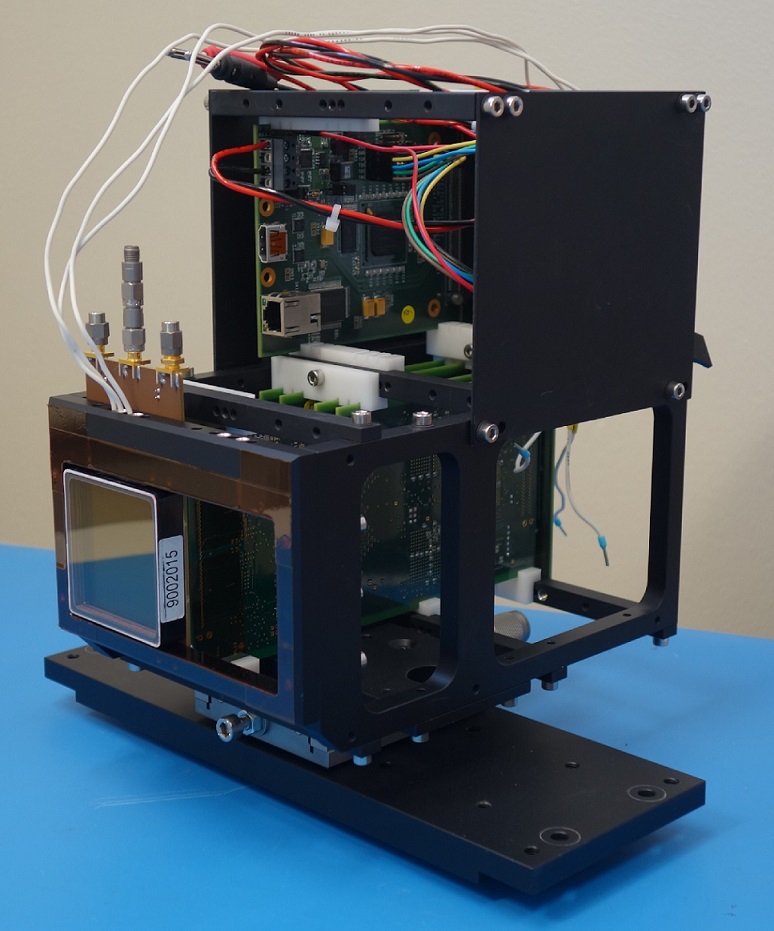}
        \end{minipage}
        \caption{Holding mechanics for the TORCH photodetector and electronics, incorporating (left) a Photek Phase-2 MCP-PMT and (right) a Photonis Planacon XP85122. The Photek tube features a square pixellated readout area embedded within a larger circular area. }
        \label{figure testbeam chariots}
        \end{center}
\end{figure}

Two independent detectors were used for the testbeam campaigns in 2015 and 2016: a single Photek Phase-2 MCP-PMT and a Photonis Planacon (model XP85122 \cite{Photonis_2014_Planacon_datasheet}), respectively. The detector assemblies can be seen in their associated holding mechanics in Figure \ref{figure testbeam chariots}. Both MCP-PMTs had their input windows spaced 0.5~mm distant from the focusing block with an air gap in between. \\

The Photek Phase-2 MCP-PMT has a 9~mm thick quartz entrance window. Whilst this is not ideal given the design of the focusing block which was fabricated with the expectation of a thinner window, the added distance can be corrected for in reconstruction. Additionally, there is a defocussing effect, which has been demonstrated to be small \cite{Castillo_Garcia_2016_thesis}. Further effects found in the Phase-2 MCP-PMT were a degraded quantum efficiency and non-homogeneity in the connection of the detector to the readout. Whilst these effects were detrimental to the photon-counting performance of the tube, they were not ultimately problematic for its operation. \\

Both the Photek tube and the Photonis tube feature an array of 32$\times$32 pixels, the latter array contained within four times the area of the former. In order to closely match the granularity requirement of TORCH, pixels were electronically connected in the horizontal direction using a mating board, in groups of eight for the Photek Phase-2 MCP-PMT and in groups of four for the Planacon, each group defining a single readout channel. For the Planacon XP85122, some difficulties were encountered in fabricating a connection between the MCP-PMT and the electronics. This meant that complete data were only obtained from four out of the eight columns of pixels. The data analysis was therefore restricted to this area. These are denoted columns 0--3 in increasing x-coordinate. \\

The relevant characteristics of the MCP-PMTs are shown in Table \ref{table MCP-PMT characteristics}.

\begin{table}[!htbp]\footnotesize
\begin{center}
\begin{tabular}{| l || c | c |}
    \hline
                        &   Testbeam period 2015        &   Testbeam period 2016      \\
    \hline
    \hline
    MCP-PMT employed    &   Photek Phase-2              &   Photonis XP85122          \\    \hline
    Number of pixels    &   4$\times$32                 &   8$\times$32               \\    \hline
    Pixel size          &   6.625$\times$0.828~mm$^2$   &   6.4$\times$1.6~mm$^2$     \\    \hline
    Instrumented area   &   26.5$\times$26.5~mm$^2$     &   51.2$\times$51.2~mm$^2$   \\    \hline
    Window material     &   Quartz                      &   Sapphire                  \\    \hline
    Window thickness    &   9~mm                        &   1~mm                      \\    \hline
    Photocathode        &   Multi-alkali (S20)          &   Bi-alkali                 \\    \hline
    Window-MCP gap      &   0.2~mm                      &   4.9~mm                    \\    \hline
    Operated gain       &   1,600,000                   &   650,000                   \\    \hline
\end{tabular}
\end{center}
\caption{Characteristics of the MCP-PMTs employed in the 2015 and 2016 testbeams. The quantum efficiency curves are shown in Figure \ref{figure intro QE curves}.}
\label{table MCP-PMT characteristics}
\end{table}

\subsection{Mechanical structure}
\label{subsection Mechanical structure}

\begin{figure}[!hbt]
    \centerline{
    \includegraphics[width=0.8\columnwidth]{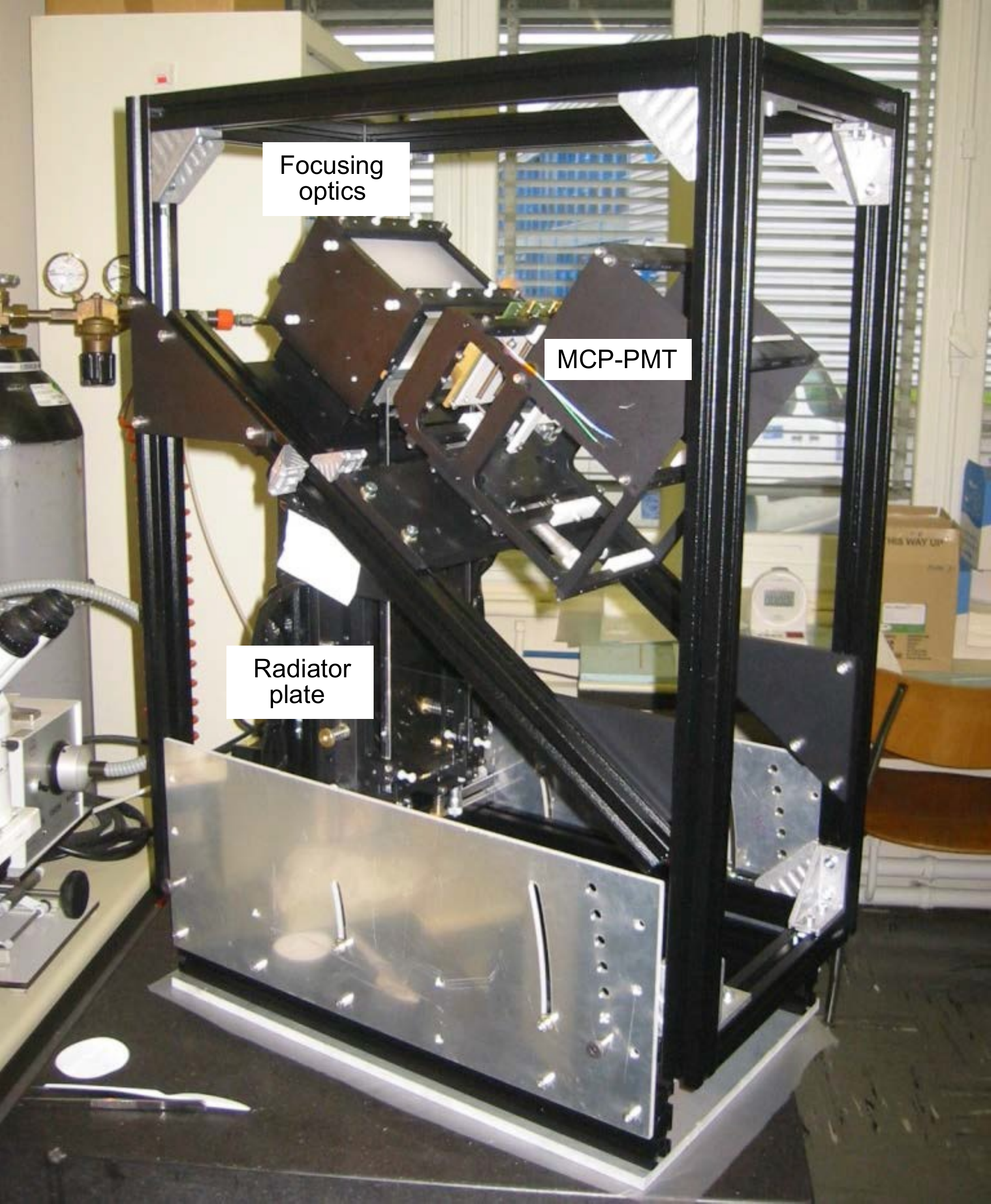}}
    \caption{The TORCH prototype module with all components mounted. The radiator, focusing block and the holding mechanics (labelled MCP-PMT) for the photodetector and electronics can be seen.}
    \label{figure PS2015 full miniTORCH}
\end{figure}

After gluing, the optical components were mounted in a rigid mechanical structure that allows rotation around the horizontal $x$-axis (perpendicular to the beam direction) to provide variation of the incident particle angle through the radiator. The structure was placed inside a light-tight box, which was then mounted on a translation table, allowing free movement in both directions perpendicular to the beam direction. A photograph of the holding mechanics, the optics, the MCP-PMT and the electronics is shown in Figure \ref{figure PS2015 full miniTORCH}. For the testbeam configuration, the full assembly was tilted at an angle of 5$^\circ$, with the top face in the downstream direction, to improve light collection from incident charged particles.

\section{The TORCH testbeam configuration}
\label{section The TORCH testbeam configuration}

Measurements were taken at the PS/T9 beam facility at CERN in 2015 and 2016, to test the TORCH prototype with positively charged particles at a nominal momentum of 5~GeV/c. A trigger system and a facility to generate high-resolution reference times for the beam particles were also deployed. Depending on the collimation and momentum settings, the charged hadron beam is mostly populated with pions and protons, with a small admixture of kaons ($\sim$1\%). \\

Two timing stations were implemented in the testbeam configuration, located approximately 10~m upstream and 1~m downstream of the TORCH prototype. Each was constructed from a bare borosilicate bar (8$\times$8$\times$100~mm$^3$) connected to a single channel MCP-PMT (Photonis PP0365G) \cite{Castillo_Garcia_2013_PP0365G_testing}. Each bar was placed in the beam at an angle close to the relevant Cherenkov angle, such that part of the generated light propagated directly towards the MCP-PMT. The corresponding signals were fed into constant fraction discriminators and transmitted via coaxial cables to the TORCH readout electronics. This gave a unified dataset incorporating both the signals from the TORCH prototype and the two time reference stations. \\

\begin{figure}[!hbt]
    \centerline
    {
    \includegraphics[width=0.75\columnwidth]{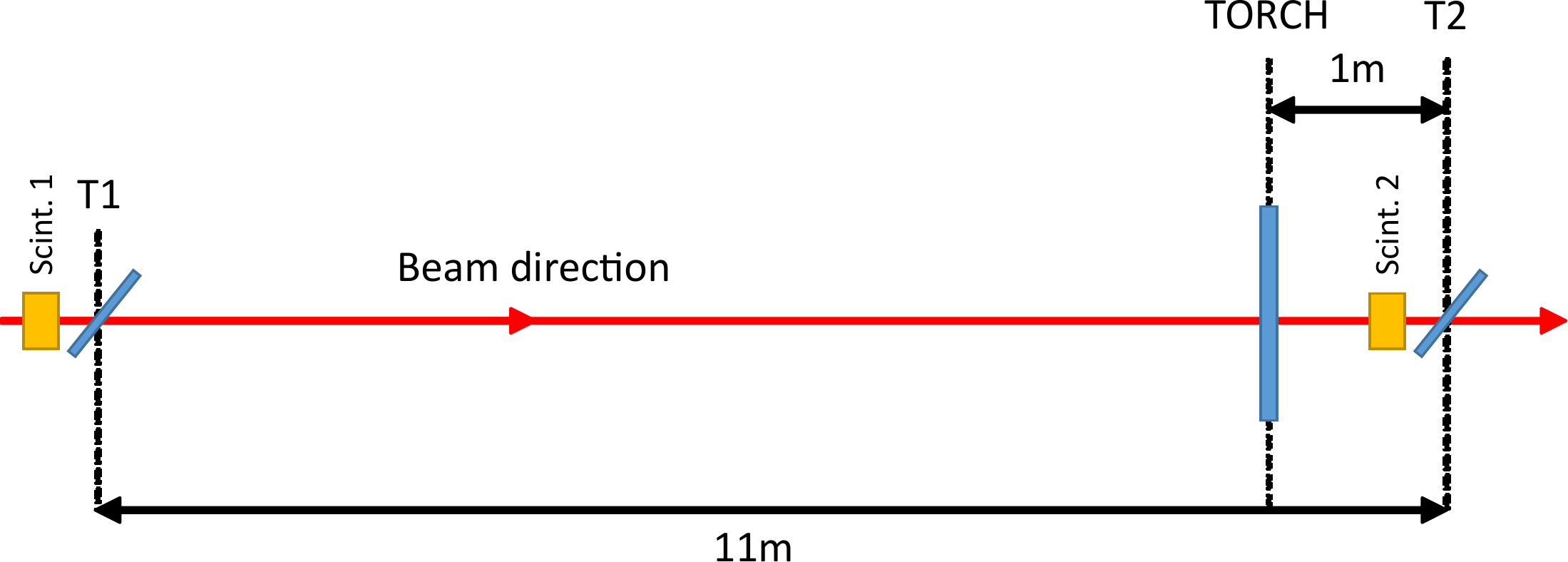}}
    \caption{Schematic overview of the beamline area showing the positioning of the timing stations T1 and T2 and the scintillators relative to the TORCH prototype. }
    \label{figure testbeam area schematic}
\end{figure}

In the 2015 period, the trigger was formed by a pair of scintillators perpendicular to the beam, each with an area of 8$\times$8~mm$^2$, and each connected to a PMT (Hamamatsu R1635-02) by a Perspex light guide. A schematic overview of the arrangement is shown in Figure \ref{figure testbeam area schematic}. The scintillators were located close to the time reference stations, ensuring incident particles passed through both borosilicate bars, and hence reducing the angular spread. However, it was found subsequently that there was a class of triggers for particles passing through the light guide, deteriorating the achievable time resolution and beam definition. Hence, in the 2016 testbeam period, this was remedied by using two scintillators in each station, with the scintillators and light guides perpendicular to each other. \\

The dual T1 and T2 time references provide redundancy of measurement and also allow for independent particle identification at relatively low momentum. Propagating over 11~m distance at 5~GeV/c, the time of flight difference between protons and pions is about 0.6~ns. The time of flight also allows determination of the momentum of the beam by measuring the average time of flight difference between pions and protons.

\section{Calibrations}
\label{section Calibrations}

To perform the data analysis, several calibrations have been incorporated: the relation of the width of a signal measured with the NINO and HPTDC to its collected charge, the Integral Non-Linearity (INL) of the HPTDC, and the time-walk of the leading edge of the input pulse due to amplitude variation. It was found during laboratory testing that the behaviour of the NINO chip is strongly dependent on the input capacitance, indicating that calibrations do not replicate from one detector to another for the 2015 and 2016 datasets. The calibrations performed will therefore be described separately for the two detectors used.

\subsection{Photek Phase-2 MCP-PMT (2015)}
\label{subsection Photek Phase-2 MCP-PMT}

The Photek Phase-2 detector was operated at an average gain of 1.6$\times$10$^6$ to ensure that the signals would be reliably detected by the electronics. As outlined in Section \ref{subsection TORCH photodetectors}, the shape of the avalanche at the anode is Gaussian, with a standard deviation of about 0.75~mm \cite{Van_Dijk_2016_Thesis, Castillo_Garcia_2016_thesis}, meaning that each single photon cluster is expected to have 3--4 hits (giving a double pulse separation of around 4~mm). \\

The position of each cluster is derived by charge-weighting the individual pixels. Due to differences found in the input capacitance between the laboratory and testbeam setups, pre-calibrations could not be used directly. The best available laboratory calibration was initially used, and the resulting charge recorded on each pixel was multiplied by a scaling factor to set the average observed charge to the same value for each channel. Based on the gain of the tube and knowledge of the charge distribution, this value was set to 80~fC.\\

The correction for INL is solely dependent on the HPTDC chips used and is expected to remain constant over time \cite{Schambach_2003_HPTDC_at_STAR}.  The contribution to the time resolution from INL, for individual signals, can be as high as 100~ps. The correction can be calculated using a dataset with high statistics. Unfortunately, the datasets available for the 2015 testbeam period were not large enough to perform this calibration reliably. Therefore correction for INL was only performed on the 2016 data. \\

The most significant correction stems from time-walk. The TORCH electronics discriminates signals of varying size using a fixed threshold. The effect of time-walk is directly correlated to the size of the signal; the smaller a signal, the longer it will take to cross the threshold, even up to a nanosecond. The correction for time-walk is made on a per-channel basis before the hits making up a cluster are combined. The assumption is that within a cluster, the time difference between hits is zero, since the individual signals represent various fractions of the same avalanche. Since it is known that the time-walk only varies with the amplitude of the input signal (here represented by the signal width), the time difference between any pair of hits within a cluster is a measure of the relative time-walk between two channels. For a given combination of two channels, the average relative time-walk can be computed as a function of the signal width measured in those two channels. Parameterizing this three-dimensional distribution then allows a derivation of the shape of the time-walk distribution for individual channels as a function of the width measured in that channel. \\

\begin{figure}
    \begin{center}
        \begin{minipage}[!hbt]{0.47\columnwidth}
            \includegraphics[width=\textwidth]{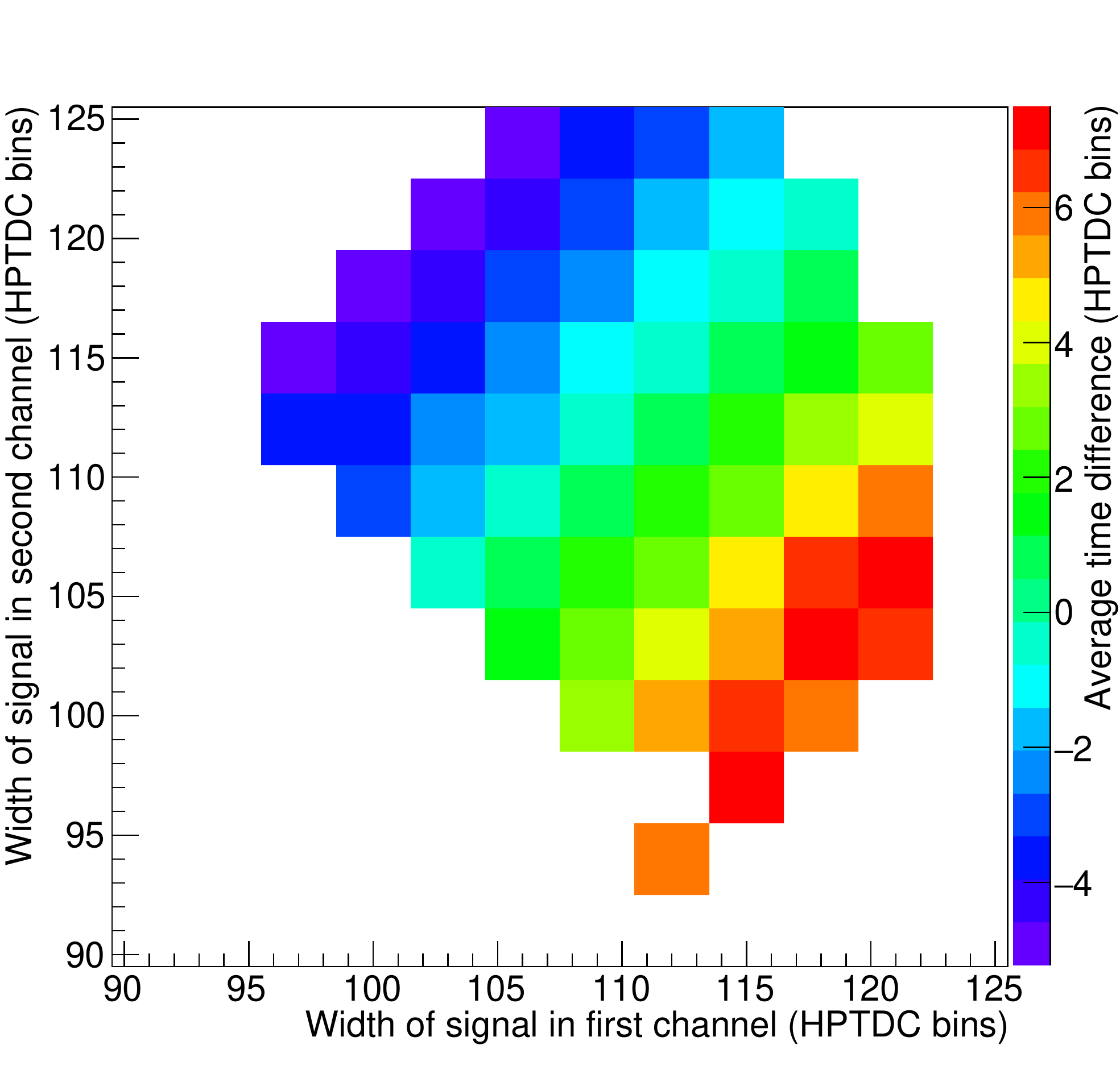}
        \end{minipage}
        \hspace{0.1cm}
        \begin{minipage}[!hbt]{0.47\columnwidth}
            \includegraphics[width=\textwidth]{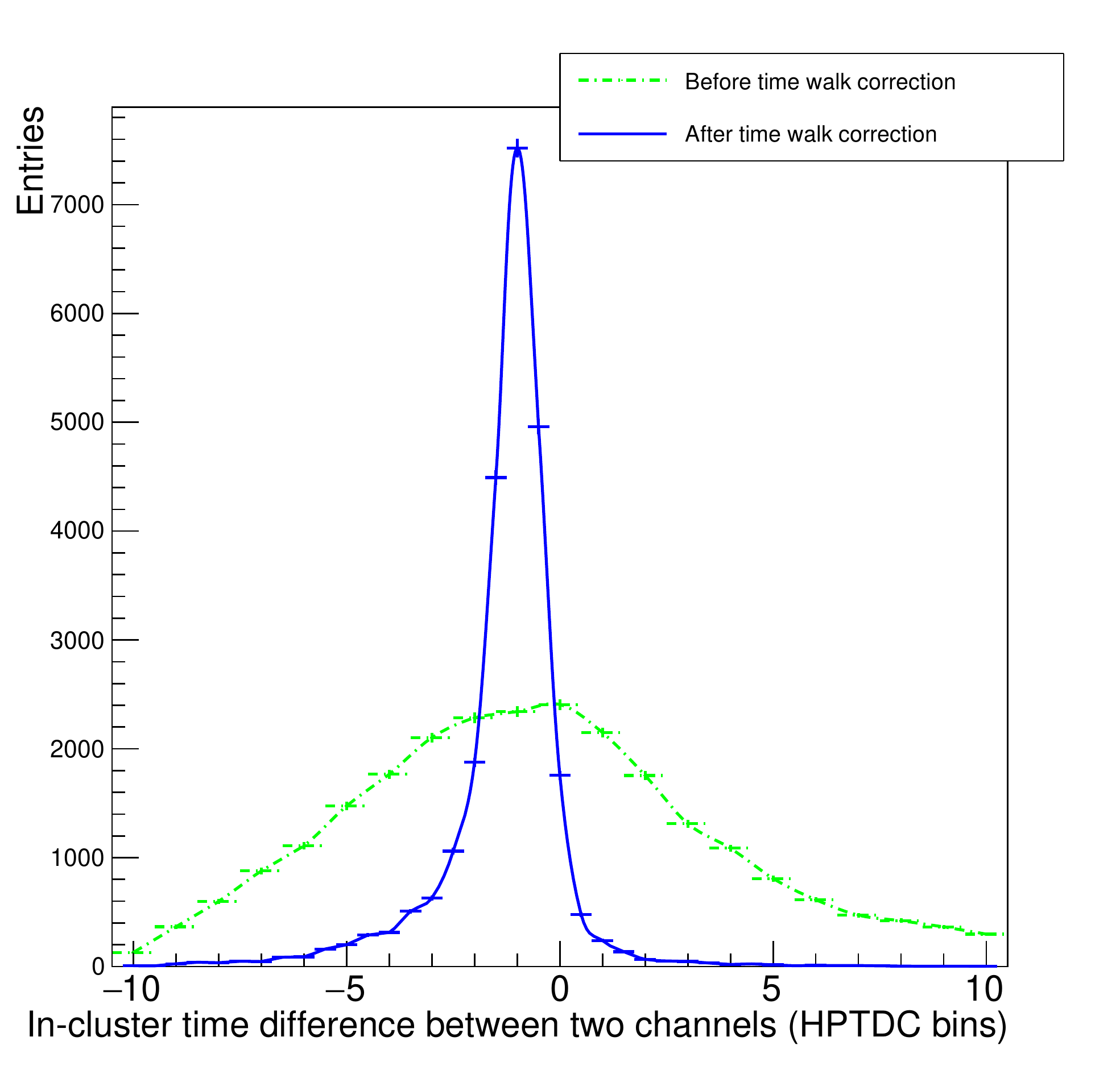}
        \end{minipage}
        \caption{(Left) average time difference between hits associated to the same cluster for a single set of two next-to-nearest-neighbour channels, expressed as a function of the width of the signal in both channels (1~HPTDC bin = 97.7~ps). The correlation between the two is extracted and fitted. (Right) Histogram of the time difference observed between the same two next-to-nearest neighbour channels, before (green) and after (blue) applying the computed time-walk correction. }
        \label{figure PS2015 time-walk demo}
        \end{center}
\end{figure}

The simplest choice is to derive the time-walk distributions from neighbours in the finely-pixellated vertical direction, since these have the largest chance of simultaneously being present in a cluster. However, in the case of the Photek Phase-2 MCP-PMT, the behaviour of the relative time difference is influenced by effects that derive from the coupling board between the MCP-PMT and the electronics. It was found during laboratory testing that the input capacitance differs significantly between channels due to longer track-lengths and/or routing on different layers. This implies that neighbouring pixels cannot be used directly to derive the time-walk correction since they show systematically different behaviour. Because of the large average cluster-size (3--4 hits), the option is available to use next-to-nearest neighbours, which electronically have similar behaviour. For each pair of these, the relative time difference distribution as a function of the width in both channels is created and then fitted; the time-walk correction on an individual channel is then derived from its correlation with the other channels. By way of an example, the average time difference as a correlated quantity between hits in two next-to-nearest neighbours is shown in Figure \ref{figure PS2015 time-walk demo}. Additionally, a histogram is shown of the time difference between next-to-nearest hits in clusters before and after applying the derived correction, for a single next-to-nearest neighbouring pair.\\

Since each channel has two next-to-nearest neighbours, the time-walk correction is improved further by averaging the fits from both sides. Finally, static offsets between channels (for example, caused by differing track-lengths) are corrected for. These are found by taking the mean average time offset between two channels after applying the relative time-walk correction.

\subsection{Photonis Planacon MCP-PMT (2016)}
\label{subsection Photonis Planacon MCP-PMT}

The individual pixel pads for the Planacon are close to twice as large in the vertical direction compared to the Photek Phase-2 detector. It is assumed the size of the avalanches from both detectors are similar, hence it is expected that each photon cluster recorded with the Planacon will have 1--2 hits. With relatively small clusters, and a high contribution from single hit clusters, the added benefit from charge-weighting the positions of the pixels in a cluster is not as significant as for the Photek Phase-2. Hence for the 2016 dataset, the choice was made not to perform charge calibration, and for multi-hit clusters the position was simply averaged. The Planacon was operated at an average gain of 6.5$\times$10$^5$. \\

During the 2016 testbeam, a high statistics dataset was recorded and subsequently used for deriving both the INL and time-walk corrections. In the construction of the mating board between the Planacon and the electronics, special care was taken to make every channel as similar as possible (in contrast to the Photek Phase-2 mating board). The time-walk calibration was performed on neighbouring channels using the same method as described above.

\section{Data analysis}
\label{section Data analysis}

For both the 2015 and 2016 datasets, the beam (as defined by the scintillator trigger) was focused very close to one of the vertical sides of the radiator. This meant that the path of light reflecting off that side differed only minimally from the light propagating directly to the MCP-PMT. As such, the number of possibilities for paths taken by the photons through the radiator reduces by a factor of two in this configuration. In the vertical direction, the beam was focused slightly below the centre of the radiator plate, at 6.4cm below (2015) and 3.6cm below (2016), respectively.

\subsection{Clustering}
\label{subsection Clustering}

\begin{figure}[!hbt]
    \centerline{
    \includegraphics[width=0.75\columnwidth]{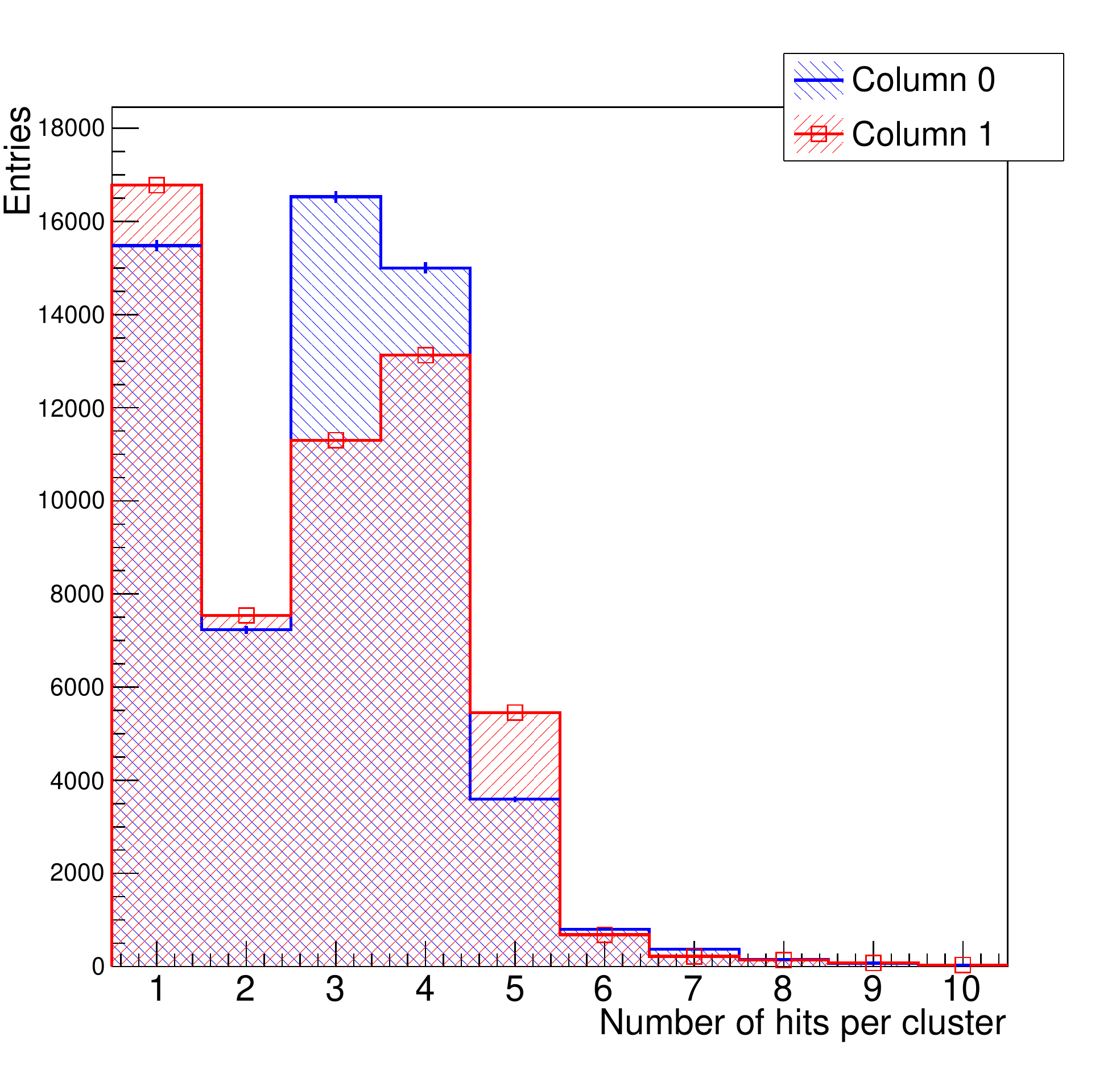}}
    \caption{Distribution of cluster sizes observed in the 2015 dataset (Photek Phase-2 MCP-PMT) for pixel columns 0 and 1. The large number of single hit clusters (relative to the expected Poisson distribution) is attributed to hits not correctly associated to a cluster, and accordingly single-hit clusters are suppressed in further data analysis.}
    \label{figure PS2015 clustersize}
\end{figure}

The clustering algorithm associates hits within a vertical column of pixels which are close in time and space.  This is defined to be within 2.5~HPTDC time bins (each 97.7~ps) after applying calibrations, and missing at most a single pixel. The columns are numbered from zero, from negative to positive horizontal detector coordinate. Clustering is not performed in the horizontal coarse pixel direction. In both the 2015 and 2016 datasets, four columns of 32~pixels are analysed. The distribution of cluster sizes for two columns in the 2015 dataset for the Photek Phase-2 MCP-PMT is shown in Figure \ref{figure PS2015 clustersize}, and is expected to follow a Poisson distribution. It can be seen that single-hit clusters are significantly enhanced. These are hits that have not been associated to the correct cluster, are an incomplete cluster, or are simply noise, and are suppressed in further data analysis. In the 2016 dataset, the cluster size of the Planacon is on average about 1.3, and all clusters are accepted.\\

Two different methods for calculating the timestamp and position of a cluster are used. For the 2015 dataset, the weighted charges from individual pixels are used to make the best possible position estimate of the true photon hit. The cluster time is further improved by charge-weighting the individual timestamps, to account for the poorer time resolution of signals with lower charge. In the case of the 2016 dataset, the position and timestamps of the pixels are simply averaged. Cluster counting measurements will be discussed in Section \ref{subsection Photon counting}.

\subsection{Particle identification}
\label{subsection Particle Identification}

\begin{figure}
    \begin{center}
        \begin{minipage}[!hbt]{0.47\columnwidth}
            \includegraphics[width=\textwidth]{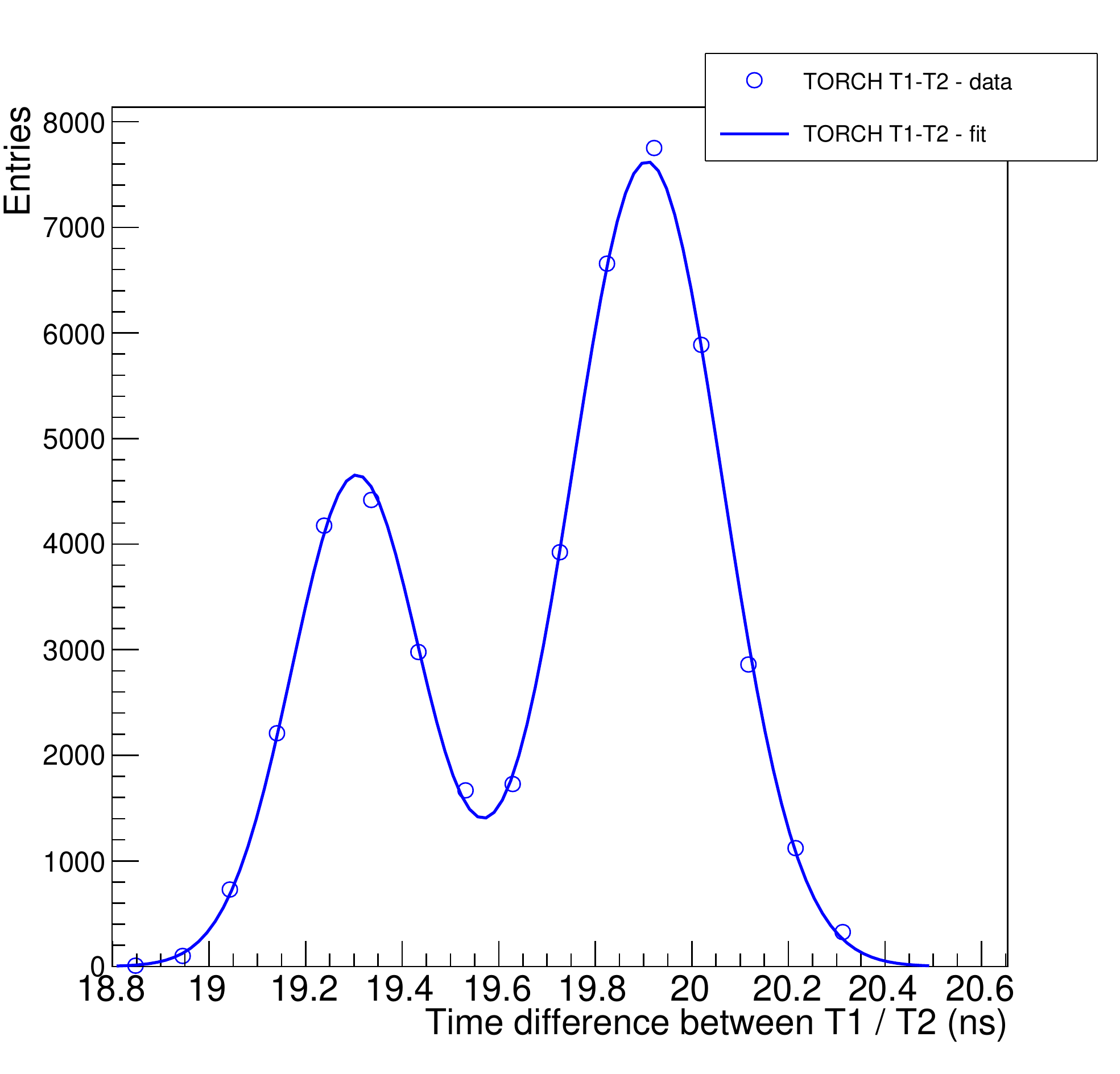}
        \end{minipage}
        \hspace{0.1cm}
        \begin{minipage}[!hbt]{0.47\columnwidth}
            \includegraphics[width=\textwidth]{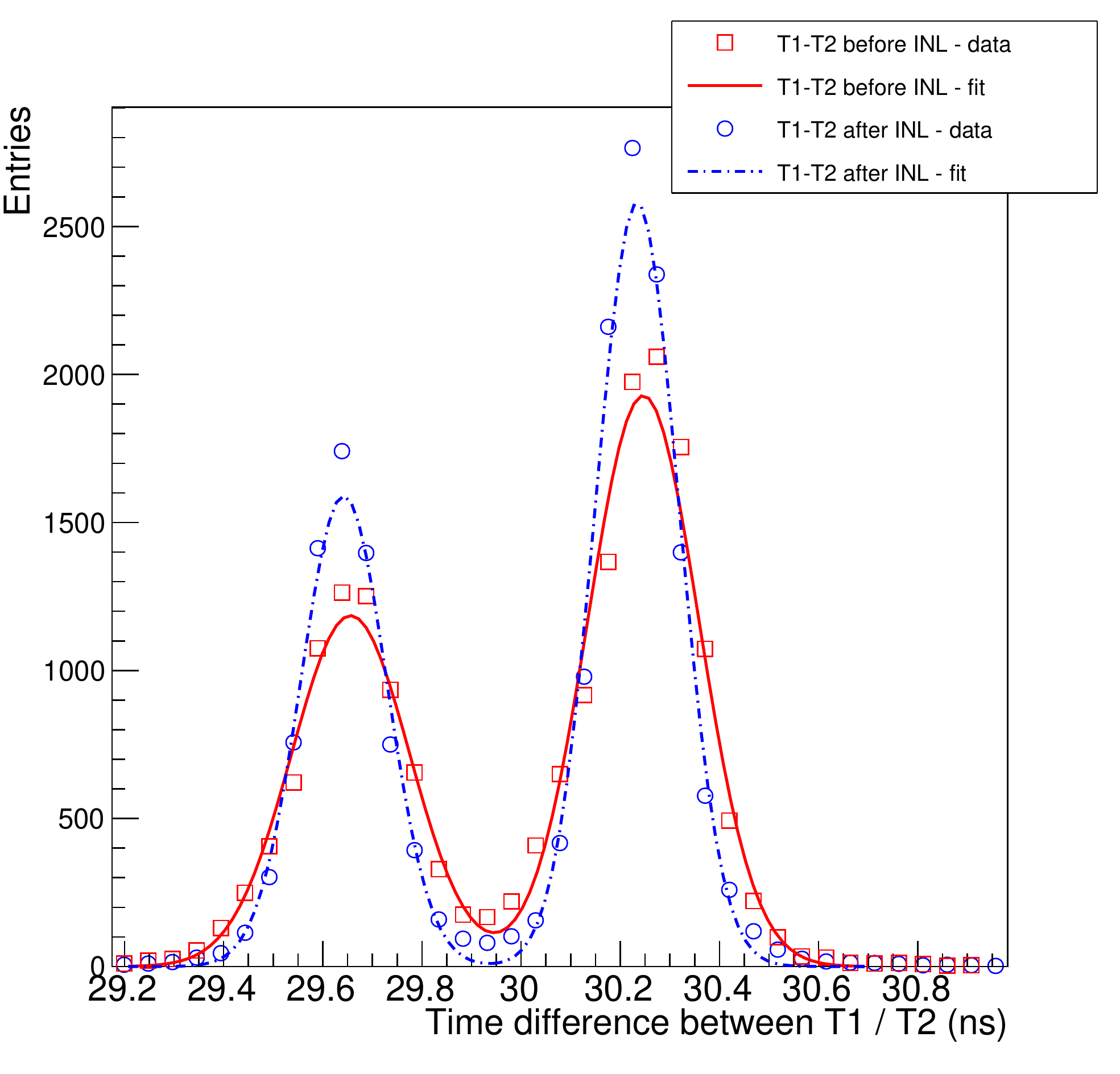}
        \end{minipage}
        \caption{Time difference measured between time reference stations T1 and T2 in 2015 (left) and in 2016 before (right, red) and after INL correction (right, blue), showing the proton peak (left in both plots) and the pion peak (right in both plots).}
        \label{figure T1-T2 particle identification}
        \end{center}
\end{figure}

The time of flight difference between the T1 and T2 stations measured with the TORCH electronics for both the 2015 and 2016 data are shown in Figure \ref{figure T1-T2 particle identification}, with the pion and proton peaks clearly seen. The proton peak, in terms of the time T1 minus T2, arrives earlier than the pion peak. This is a consequence of a long cable for T1 and a short cable for T2 effectively inverting the underlying distribution in time. The standard deviations of the fitted data are given in Table \ref{table T1-T2 PID sigmas} and demonstrate the combined quality of the time reference signals. It is expected that there is also a small admixture of kaons (about $\sim$1\%), however this contribution cannot be distinguished in either case. The figures for 2016 show that the INL correction is performed to good effect, significantly reducing the time spread.\\

\begin{table}[!htbp]\footnotesize
\begin{center}
\begin{tabular}{| l || l | l |}
    \hline
                        &   Proton peak         &   Pion peak       \\
    \hline
    \hline
    2015                &   134.0$\pm$0.9~ps    &   156.0$\pm$0.9~ps \\  \hline
    2016 (before INL)   &   119$\pm$1~ps        &   112$\pm$1~ps    \\  \hline
    2016 (after INL)    &   87$\pm$1~ps         &   84.0$\pm$0.7~ps \\  \hline
\end{tabular}
\end{center}
\caption{Standard deviations of fits to the data shown in Figure \ref{figure T1-T2 particle identification}.}
\label{table T1-T2 PID sigmas}
\end{table}

In 2015, the time of flight difference measured between pions and protons is 601$\pm$2~ps, with the uncertainty derived from the error on the means of the Gaussian fits to the data. From the time of flight difference, a momentum of 5.14$\pm$0.01~GeV/c is calculated, deviating slightly from the nominal beam settings (5~GeV/c). In 2016, the time of flight measured between pions and protons is 592$\pm$2~ps, giving a momentum of 5.18$\pm$0.01~GeV/c, again deviating slightly from nominal.

\subsection{Timing performance}
\label{subsection Timing performance}

The position and timestamp for a given cluster are calculated after applying the calibrations to the data. The clusters are then separated into pion and proton contributions according to the T1-T2 time of flight (Figure \ref{figure T1-T2 particle identification}). The ambiguous regions between the proton and pion peaks (in 2015 set to 19.4--19.75 ns, in 2016 set to 29.8--30.1 ns) are removed from further analysis. \\

\begin{figure}
    \begin{center}
        \begin{minipage}[!hbt]{0.47\columnwidth}
            \includegraphics[width=\textwidth]{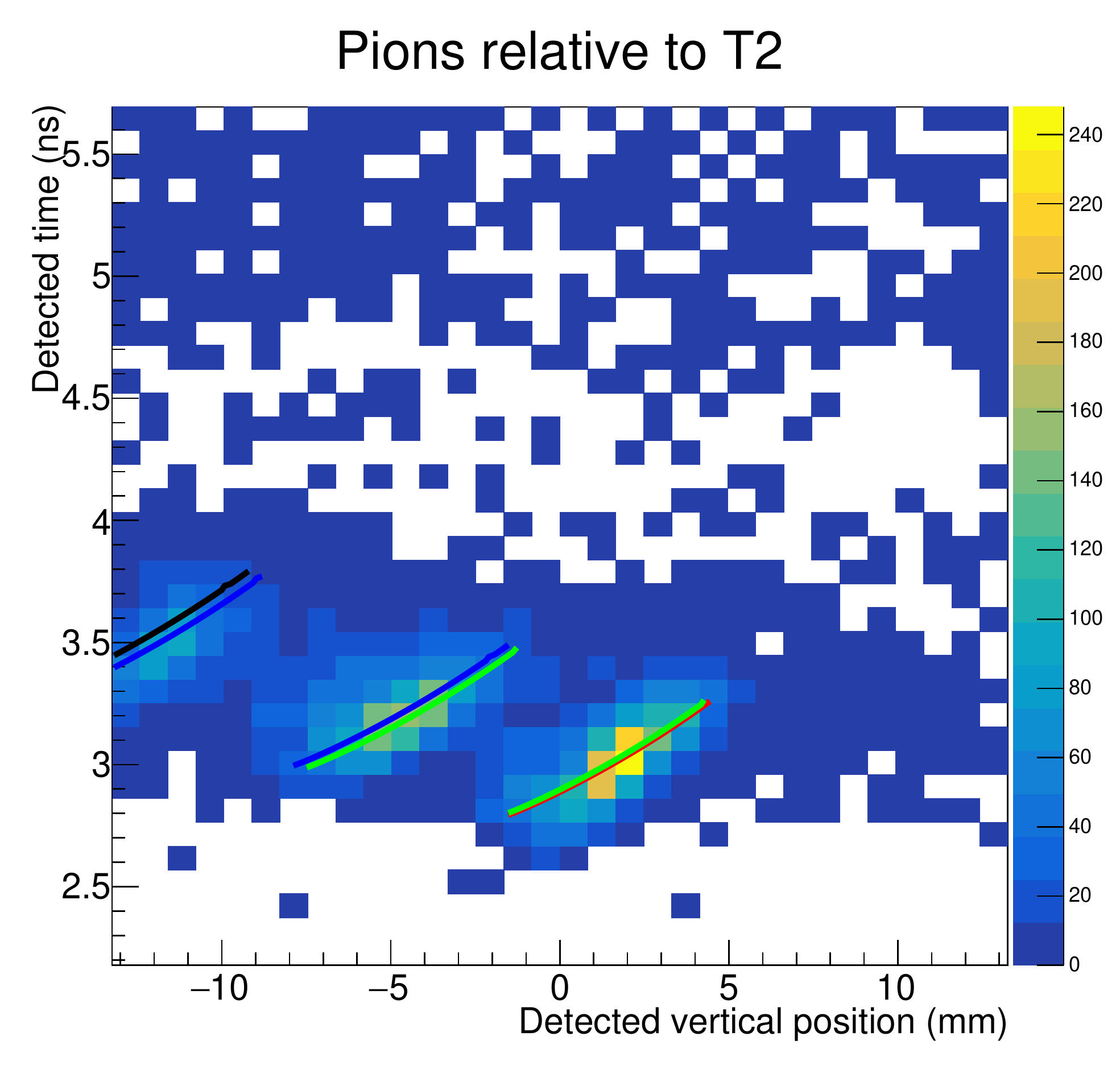}
        \end{minipage}
        \hspace{0.1cm}
        \begin{minipage}[!hbt]{0.47\columnwidth}
            \includegraphics[width=\textwidth]{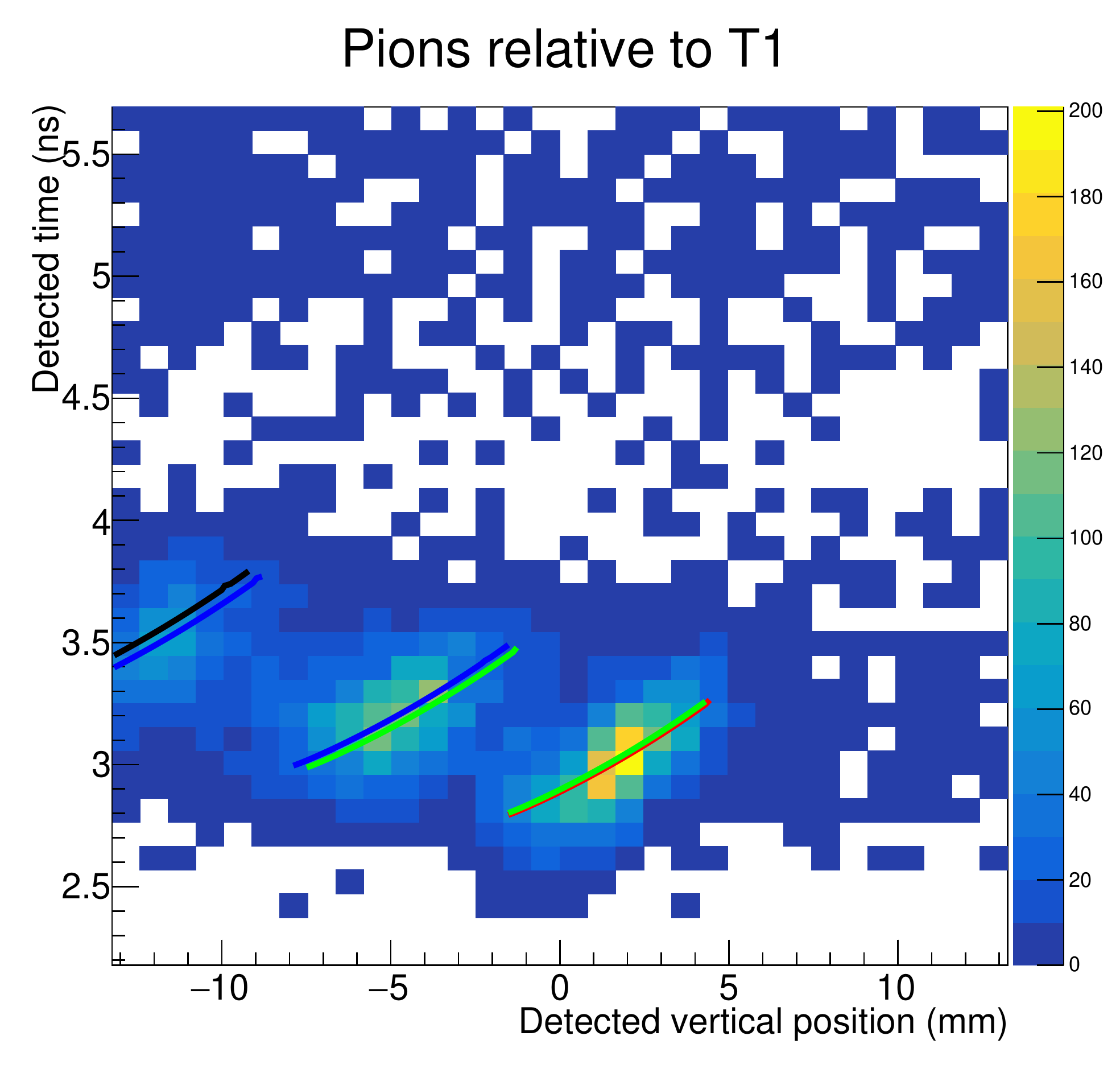}
         \end{minipage}
    \end{center}
    \vspace{0.1mm}
    \begin{center}
        \begin{minipage}[!hbt]{0.47\columnwidth}
            \includegraphics[width=\textwidth]{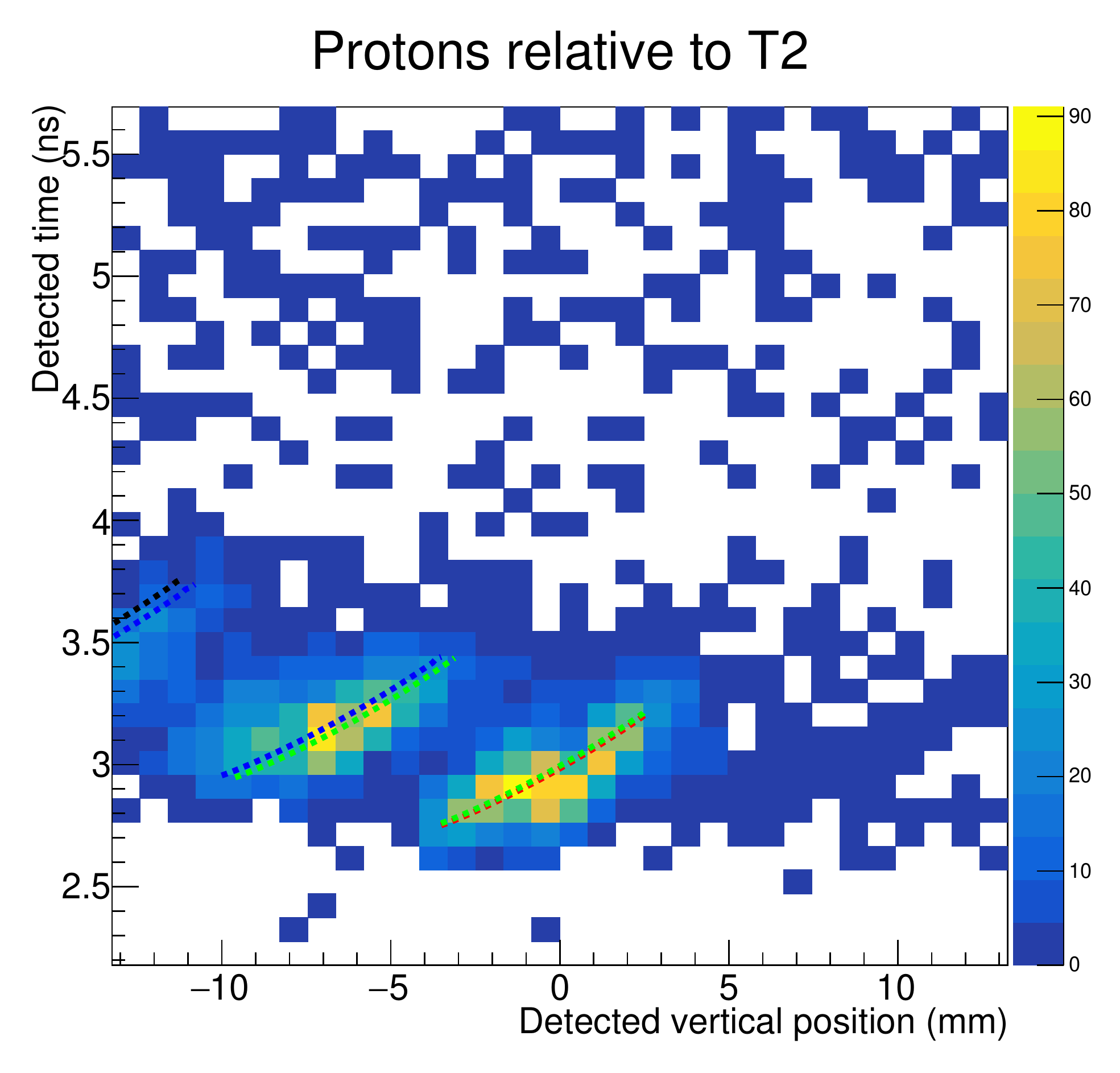}
        \end{minipage}
        \hspace{0.1cm}
        \begin{minipage}[!hbt]{0.47\columnwidth}
            \includegraphics[width=\textwidth]{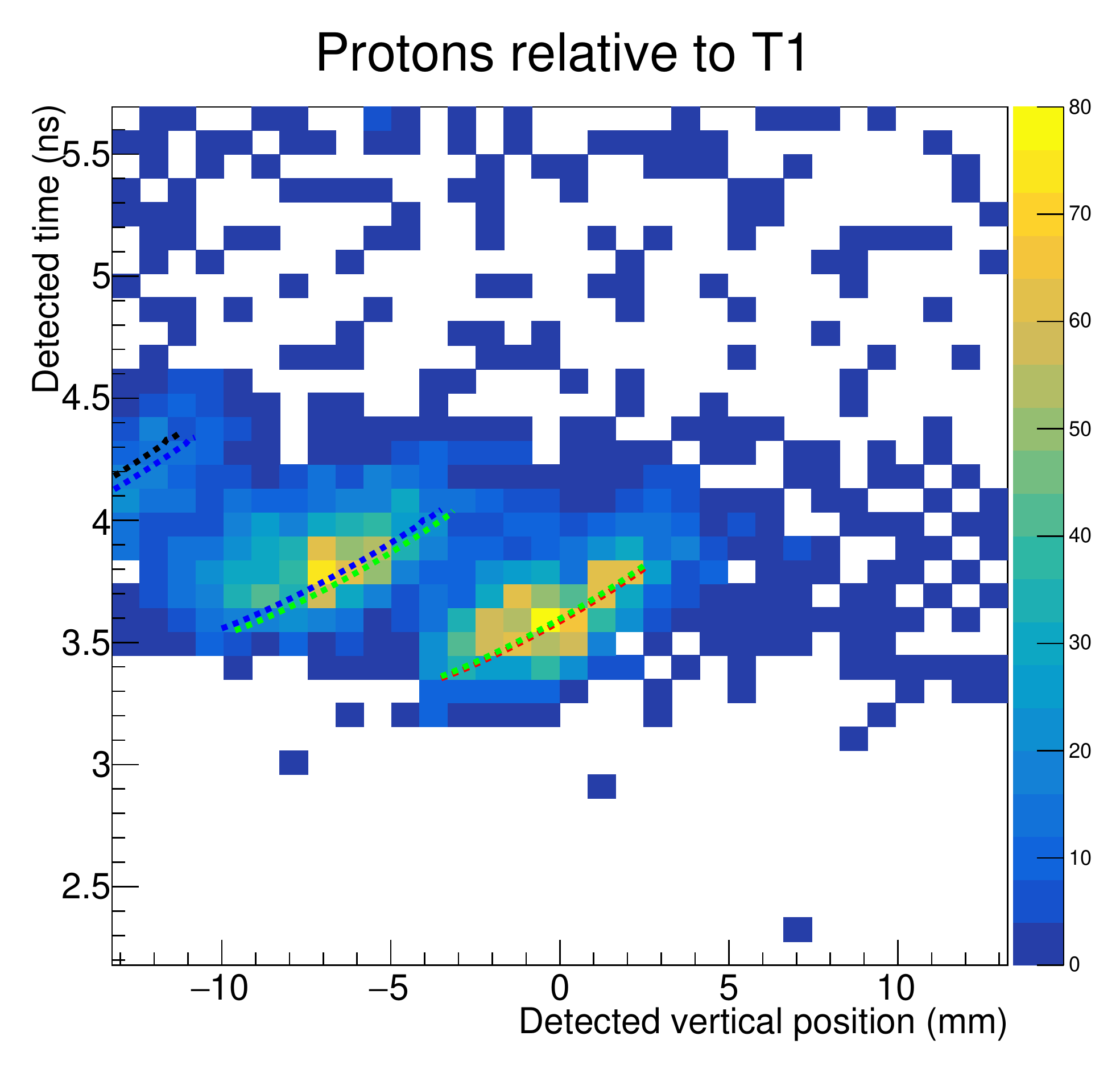}
        \end{minipage}
        \caption{Data taken in 2015: detected vertical position versus timestamp for clusters detected in column 0, after selecting for pions (top) and protons (bottom), relative to timing signal T2 (left) and T1 (right). The overlaid lines represent the simulated patterns for direct light (red) and light undergoing a single (green), double (blue) or triple (black) reflection off the vertical side faces of the optics (see Figure \ref{figure intro double side reflection}). }
         \label{figure PS2015 time projections}
    \end{center}
\end{figure}

\begin{figure}[!hbt]
    \centerline{
    \includegraphics[width=0.75\columnwidth]{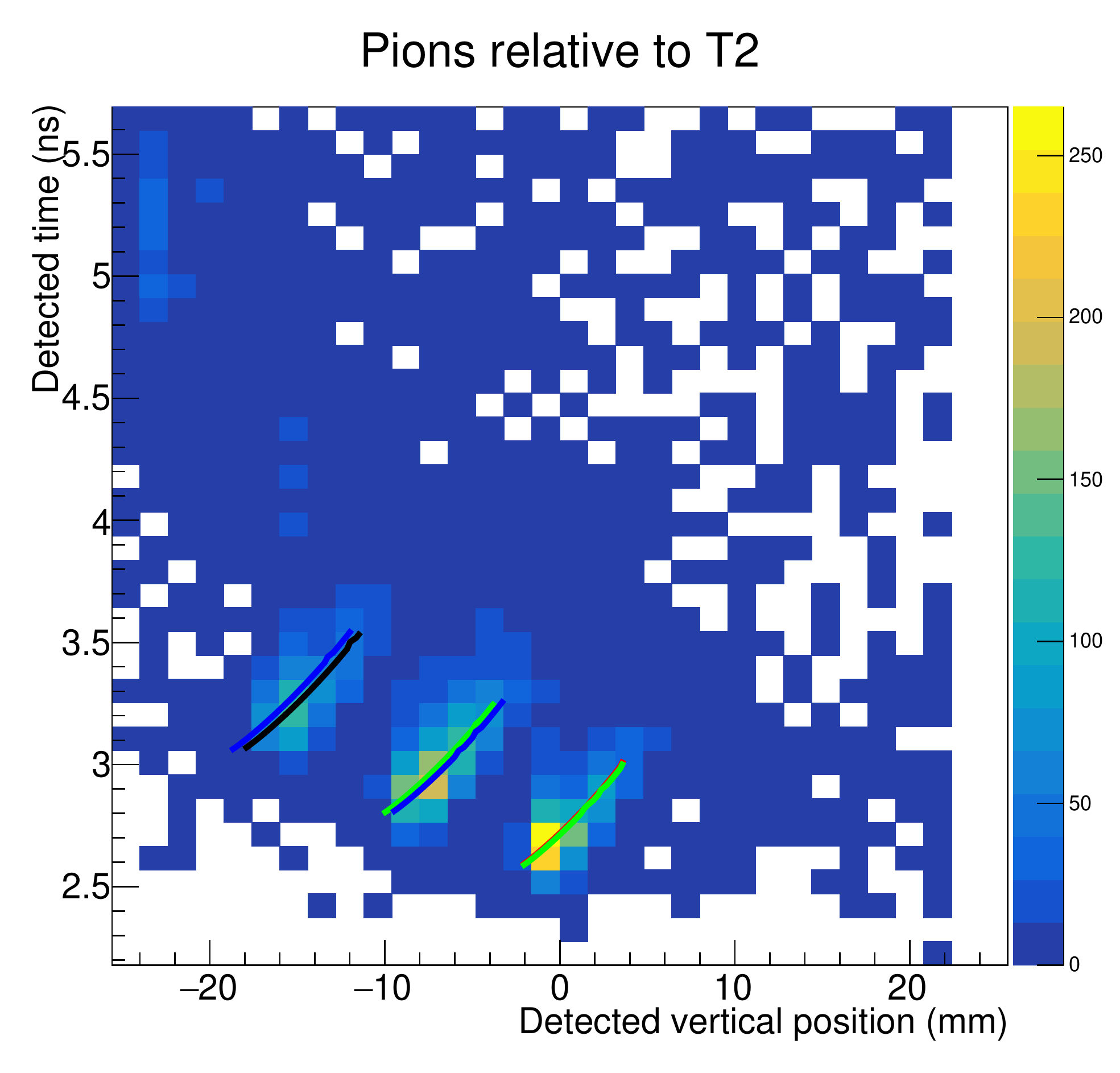}}
    \caption{Pion-selected data taken in 2016: detected vertical position versus timestamp for clusters detected in column 1 relative to time reference station T2. The overlaid lines represent the simulated patterns for direct light (red) and light undergoing a single (green), double (blue) or triple (black) reflection off the vertical sides of the optics. }
    \label{figure PS2016 time projections}
\end{figure}

The time relative to T1 and T2 as a function of measured vertical (finely-pixellated) position for clusters detected in column 0 of the Photek MCP-PMT in the 2015 dataset is shown in Figure \ref{figure PS2015 time projections}. The expected patterns from simulation are overlaid. The pattern folding (see Section \ref{section Design of TORCH}) is clearly visible; multiple patterns are observed. The overlaid patterns appear in closely-spaced paired groups, both from the direct light and from pattern folding off the vertical side close to where the charged particle beam impinges on the radiator. Comparing top and bottom plots, there is a shift in the position of the patterns between pions and protons caused by the difference in the Cherenkov angles, expected to be 14.4~mrad at 5.14~GeV/c (equivalent to a shift of 2.3~pixels). There is also an observed deterioration visible in the timing resolution of the T1 plots relative to the T2 plots; this is due to signal degradation over the length of the cable transporting the T1 signal to the TORCH electronics. A slight discontinuity exists at the centre of each pixel column, when comparing Figure \ref{figure PS2015 time projections} (top, left) and (bottom, left). This position correlates to the edge of two individual NINO chips; the discontinuity indicates that constructing the time walk correction across this boundary could be further optimized. \\

Data taking was improved in several aspects in 2016. Firstly, the charged particle beam was focused on a number of different positions on the radiator plate, allowing for alignment of the detector from data. Secondly, recording of a very large dataset allowed for improvements of the calibration, especially the INL. \\

Figure \ref{figure PS2016 time projections} shows the MCP-PMT time measurement relative to T2, detected on column 1 of the Planacon, as a function of vertical position for selected pions. As in the 2015 testbeam period, the observed pattern closely agrees with Monte Carlo expectations. Comparing Figures \ref{figure PS2015 time projections} and \ref{figure PS2016 time projections}, it should be remembered that the vertical pixel width of the Planacon is twice as large as that of the Photek Phase-2. Also the overall size of the Planacon detector is vertically twice as large. \\

\begin{figure}
    \begin{center}
        \begin{minipage}[!hbt]{0.47\columnwidth}
            \includegraphics[width=\textwidth]{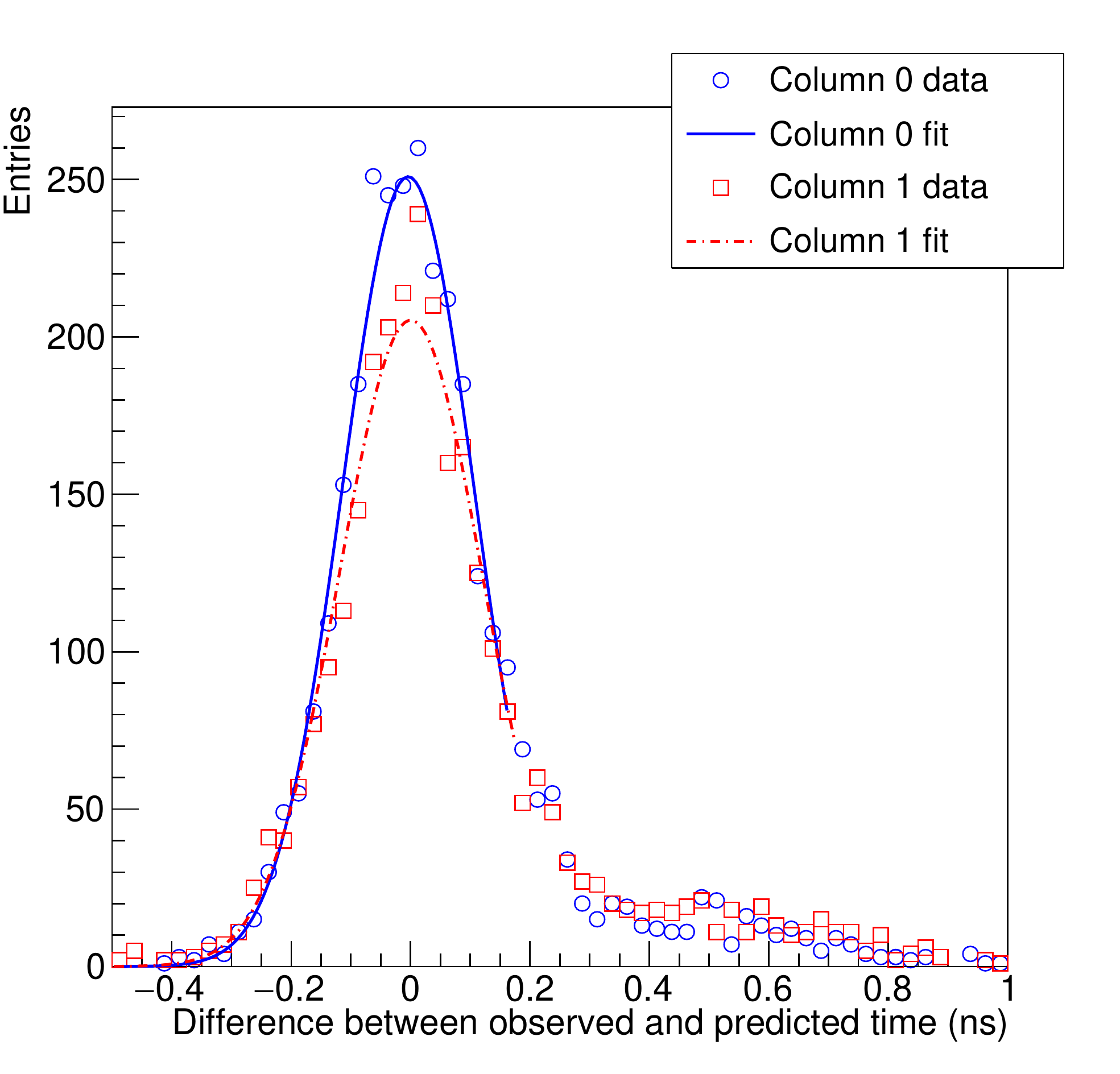}
        \end{minipage}
        \hspace{0.1cm}
        \begin{minipage}[!hbt]{0.47\columnwidth}
            \includegraphics[width=\textwidth]{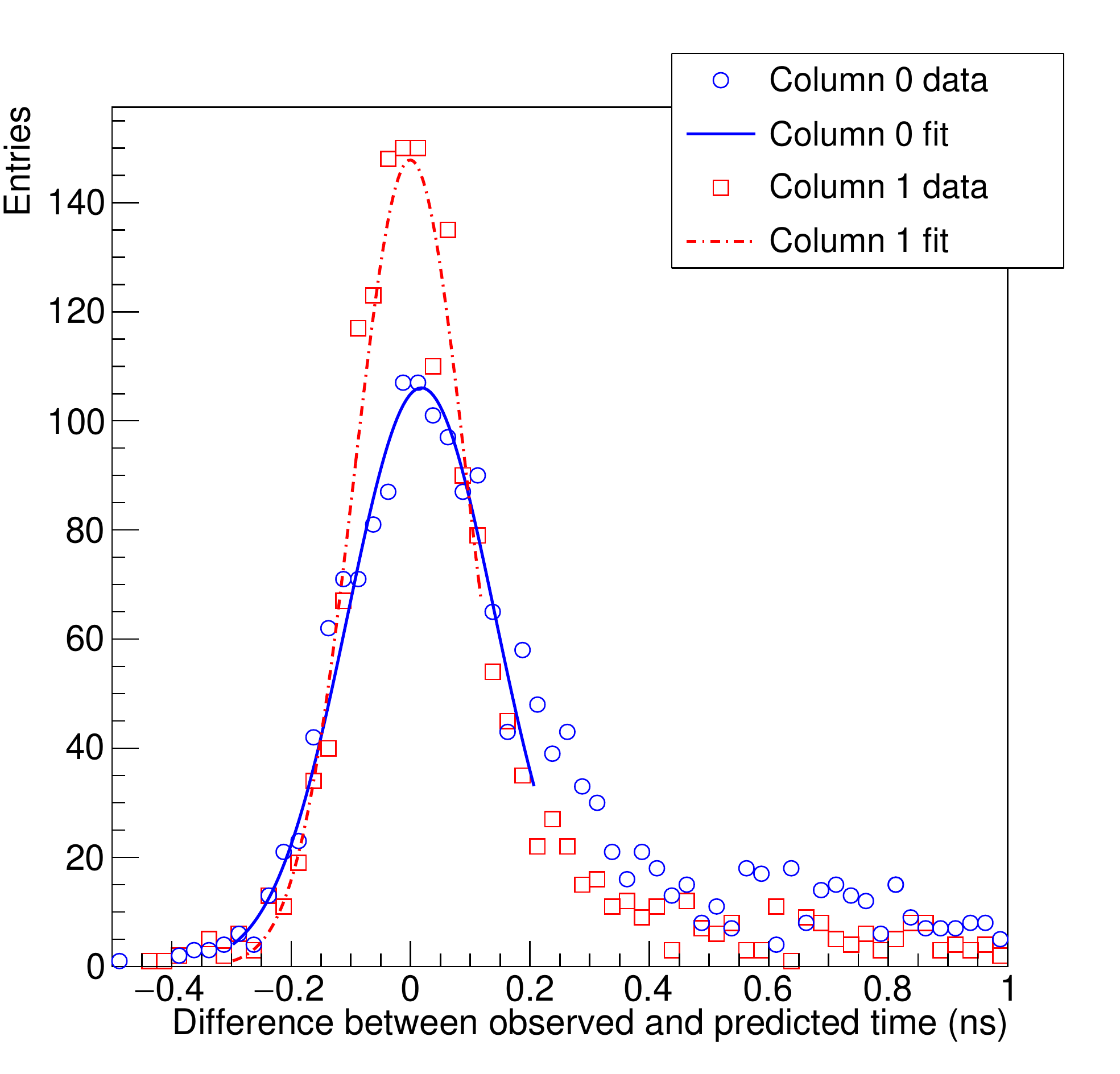}
        \end{minipage}
        \caption{Difference between observed and predicted times for two out of four columns deployed in 2015 (left) and in 2016 (right), for prompt photons from pions. The tail to the right is attributed to non-optimal time-walk corrections (left) and backscattered photo-electrons (right).}
        \label{figure td-tdr}
        \end{center}
\end{figure}

The prompt part of the pion signal relative to time reference station T2 (Figure \ref{figure PS2015 time projections} (top, left) and Figure \ref{figure PS2016 time projections}) is now used to benchmark the timing performance. For this measurement the prompt part of the pattern is used, composed of light with either no reflection off the vertical side face or with just a single reflection off the side close to where the charged particle beam traversed the radiator. Residuals of the measured times relative to the predicted curves are shown for two columns of the Photek Phase-2 tube and the Planacon in Figure \ref{figure td-tdr}. Table \ref{table td-tdr results} lists the standard deviations of Gaussian fits to these timing residuals, which is indicative of the single-photon timing resolution achieved. Note that variation due to smearing from the time reference station has not yet been corrected for. The gap between the entrance window and the first micro-channel plate of the Photek Phase-2 MCP-PMT is small (0.2~mm), hence the tail at later times seen in Figure \ref{figure td-tdr} (left) cannot be attributed to backscattering. The most likely cause is a non-optimal time-walk correction. In the case of the Planacon, the input gap is large (4.9~mm), which is expected to displace the backscattering peak to about 0.5--1~ns after the main peak. Hence the tail in Figure \ref{figure td-tdr} (right) is attributed to backscattering. Despite the much coarser granularity of the Planacon, the timing performance has been maintained in that particular TORCH configuration, mainly due to improvements in the calibration techniques. \\

\begin{table}[!htbp]\footnotesize
\begin{center}
\begin{tabular}{| l || l | l |}
    \hline
                &   2015            &   2016                \\
                &   $\sigma$ of fit &   $\sigma$ of fit     \\
    \hline
    \hline
    Column 0    &   110$\pm$2~ps    &   124$\pm$4~ps        \\  \hline
    Column 1    &   120$\pm$3~ps    &   94$\pm$3~ps         \\  \hline
    Column 2    &   137$\pm$3~ps    &   103$\pm$3~ps        \\  \hline
    Column 3    &   111$\pm$3~ps    &   99$\pm$4~ps         \\  \hline
\end{tabular}
\end{center}
\caption{Standard deviation of Gaussian fits to the timing residuals for all columns, for 2015 and 2016 datasets.}
\label{table td-tdr results}
\end{table}

To derive the intrinsic single-photon time resolution of TORCH, an estimate needs to be made of the time resolution of the time reference stations. As T1 and T2 have identical construction, it is assumed that the time resolution is the same, but that T1 suffers extra smearing from the long cable over which the signal is propagated. In 2015, insufficient data were collected to perform this subtraction reliably. However in 2016 data, the contribution from signal propagation can be factored out, and is found to be 56$\pm$14~ps, where the error is statistical. From the INL-corrected pion distribution shown in Figure \ref{figure T1-T2 particle identification} (right, blue) it is then estimated that the intrinsic time resolution of a single time reference station is 44$\pm$9~ps. Subtracting in quadrature this contribution from the resolutions quoted in Table \ref{table td-tdr results} gives a range of (83--115)$\pm$6~ps for the time resolution of the TORCH prototype. \\

\subsection{Photon counting}
\label{subsection Photon counting}

A photon counting efficiency measurement was performed only on data from the Planacon MCP-PMT. A photocathode degradation issue with the Photek Phase-2 MCP-PMT made the data for that tube less reliable. \\

The photon counting efficiency  of the TORCH prototype is calculated by comparing the number of photons detected per event to a GEANT4 simulation \cite{Geant4_2003_main_paper, Van_Dijk_2016_Thesis}. The simulation accounts for losses due to Rayleigh scattering, and a Lambertian model is used for losses due to microscopic surface roughness of reflective faces. The simulation also accounts for Fresnel reflections at the air gap between the exit surface of the focusing block and the detector window. The resulting photon spectrum is then modified using the transmission curve of the glue used between the radiator plate and the focusing block, the reflectivity of the aluminium surface of the focusing block and the quantum efficiency of Planacon (see Figures \ref{figure intro mirror reflectivity} and \ref{figure intro QE curves}). To account for the collection efficiency of the tube, an efficiency of 65\% was applied. \\

The final applied efficiency factor derives from the threshold of the NINO chip. Two types of cluster inefficiency are considered: those for which the charge measured is so low that the signal does not exceed the threshold, and those straddling the border between two pixels, dividing the charge in such a way that neither meets the threshold. Measurements had previously been performed at several NINO settings and the charge threshold was found to be between 30--60~fC. Following estimates from testbeam data, it is assumed that an average threshold level of 42~fC is representative. Assuming a Gaussian distribution with representative average gain and geometrical spread, it is then estimated that 12.7\% of the total number of generated photoelectrons are lost on average. To simulate this loss, an additional random cut is placed on the number of photoelectrons. \\

\begin{figure}[!hbt]
    \centerline{
    \includegraphics[width=0.75\columnwidth]{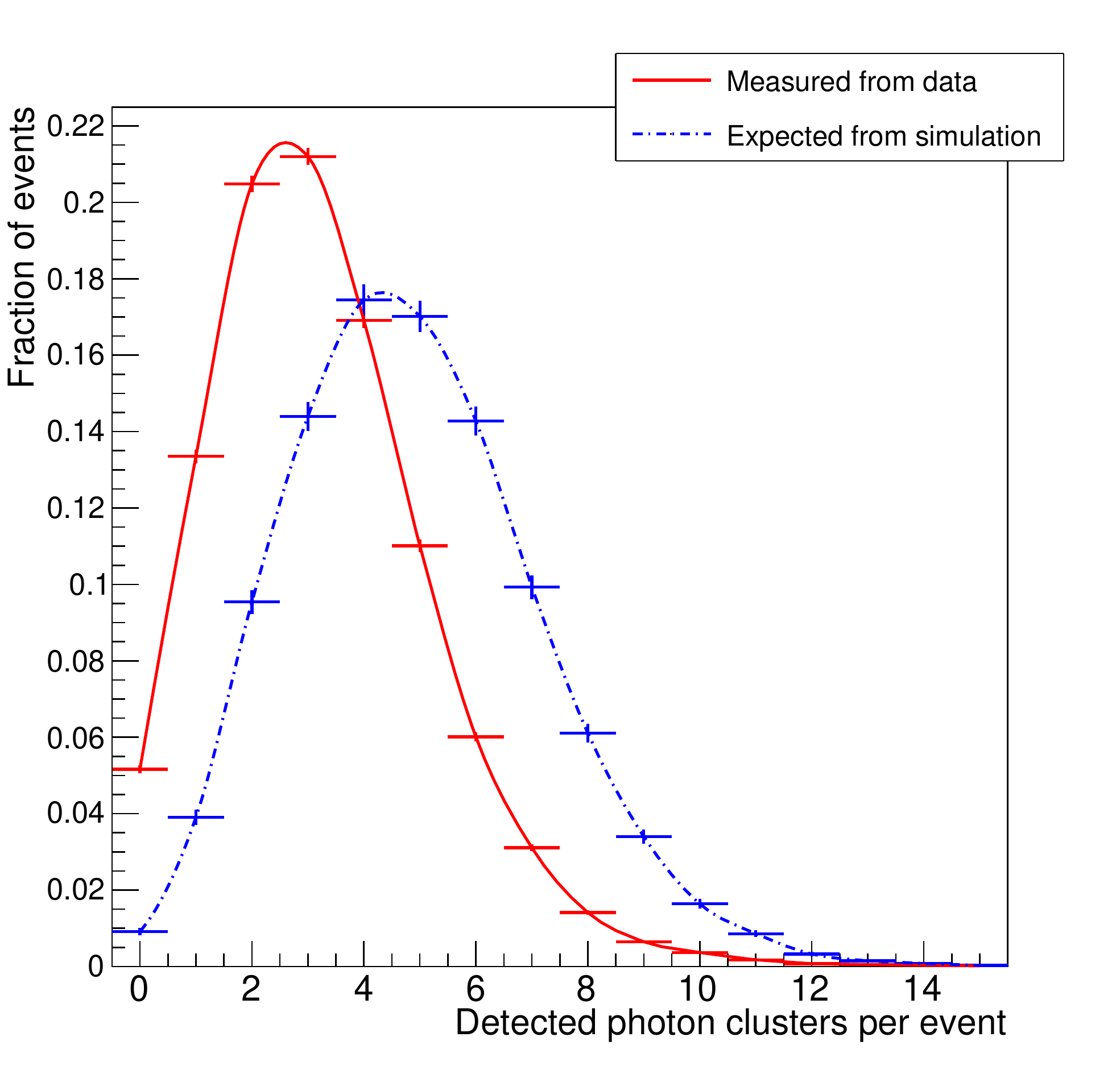}}
    \caption{Measured photon counting statistics per event in 2016 testbeam data (red) and expected from simulation (blue).}
    \label{figure PS2016 photon counting}
\end{figure}

Detector patterns for 10k protons and 10k pions were generated, with the effects described above applied. To derive the average expected number of photons over all events, the pion and proton distributions were weighted and combined according to their relative fractions from the integrals under the pion and proton distributions (see Figure \ref{figure T1-T2 particle identification}), namely 61$\pm$3\% pions and 39$\pm$2\% protons. The resulting yields from data and simulation are shown in Figure \ref{figure PS2016 photon counting}. \\

The mean number of photons expected from simulation is 4.89$\pm$0.02, compared to 3.23$\pm$0.01 observed in data (statistical errors only). Therefore, on average about 34\% fewer photons are observed than expected, indicating that additional factors remain to be accounted for. This will be studied in future developments planned for the TORCH project.

\section{Summary and future plans}
\label{section Conclusion}
TORCH is a DIRC-type detector, designed to achieve high-precision time-of-flight over large areas. In order to provide a $K-\pi$ separation up to 10\,GeV/$c$ momentum over a 10\,m flight path, a ToF resolution of $\sim$15\,ps is required. This translates to a per-photon resolution of 70\,ps, given around 30 detected photons per track.\\

A small-scale TORCH demonstrator, with a quartz plate of dimensions 120$\times$350$\times$10~mm$^3$ has been constructed. The detector is read out by a single customised Photek MCP-PMT with 32$\times$32 pixels contained within an area 26.5$\times$26.5mm$^2$ square, and where a charge-division technique has been used to improve the spatial granularity. Testbeam results are compared to those from a commercial 2-inch square Planacon 32$\times$32 pixellated MCP-PMT.\\

The data analysis methods employ a data-driven approach to correct simultaneously for time-walk, charge-to-width calibration, and integral non-linearities of the electronics readout. Following significant improvements to the triggering and calibration techniques, a range of (83--115)$\pm$6~ps is measured for the single-photon time resolution of TORCH. Hence the single-photon timing performance is approaching the required 70\,ps per photon. The single-photon counting performance is around 34\% lower than expected from simulation. Improvements in the electronics calibration techniques and threshold control are expected in the future. In conclusion, the testbeam measurements have demonstrated the principle of operation of TORCH, with a timing resolution that approaches the requirement for the final detector. \\

The small-scale demonstrator is a precursor to a full-scale TORCH module (660$\times$1250$\times$10~mm$^3$), which is currently under construction. The module will be equipped with ten full-sized 2-inch Photek Phase-3 64$\times$64 pixel MCP-PMTs. The MCP-PMTs, optics and electronics to equip this module have been delivered and are currently under test. All components, including the mechanical structure and housings, will be ready for testbeam operation in 2018.

\section*{Acknowledgements}

The support of the European Research Council is gratefully acknowledged in the funding of this work through an Advanced Grant under the Seventh Framework Programme FP7 (ERC-2011-AdG 299175-TORCH). The authors wish to express their gratitude to Simon Pyatt of the University of Birmingham for wire bonding the NINO ASICs, and the CERN EP-DT-EF and TE-MPE-EM groups for their efforts in the coupling PCB design and bonding. We are grateful to Nigel Hay, Dominic Kent and Chris Slatter of Photek for their work on the MCP-PMT development.

\end{document}